\documentclass[a4paper,10pt,twocolumn,nobalancelastpage,aps,pra,superscriptaddress,nofootinbib,longbibliography]{revtex4-1}
\usepackage{tikz}
\usepackage{graphicx}

\graphicspath{ {images/} }

\usepackage{amsmath, amsthm, amssymb, amsfonts, dsfont, bm, bbm}

\usepackage{setspace}
\def\thickhline{\noalign{\hrule height.8pt}}

\usepackage{soul} % Highlighting etc.

\usepackage[pdftex,
	colorlinks,
	allcolors=blue,
	pdfauthor={},
    pdftitle={Bayesian optimization of non-classical optomechanical correlations},
	pdfsubject={},
	pdfkeywords={},
	pdfproducer={Latex with hyperref},
	pdfcreator={pdflatex}]{hyperref}

\renewcommand{\eqref}[1]{Eq.~(\ref{#1})}
\newcommand{\figref}[1]{Fig.~\ref{#1}}
\newcommand{\tblref}[1]{Table~\ref{#1}}

\newcommand{\appxref}[1]{Appendix~\ref{#1}}
\newcommand{\secref}[1]{Section~\ref{#1}}

\usepackage{listings}
\usepackage{lipsum}

\usepackage[utf8]{inputenc}

\DeclareFixedFont{\ttb}{T1}{txtt}{bx}{n}{11} % for bold
\DeclareFixedFont{\ttm}{T1}{txtt}{m}{n}{11}  % for normal

\usepackage{color}
\definecolor{deepblue}{rgb}{0,0,0.5}
\definecolor{deepred}{rgb}{0.6,0,0}
\definecolor{deepgreen}{rgb}{0,0.5,0}

\usepackage{listings}

\newcommand\pythonstyle{\lstset{
        language=Python,
        basicstyle=\ttfamily\small,
        otherkeywords={self},             % Add keywords here
        keywordstyle=\ttb\color{deepblue},
        emph={MyClass,__init__},          % Custom highlighting
        emphstyle=\ttb\color{deepred},    % Custom highlighting style
        stringstyle=\color{deepgreen},
        frame=tb,                         % Any extra options here
        showstringspaces=false            %
}}

\newcommand{\iu}{{\mathrm{i}\mkern1mu}}

\newcommand{\me}{\mathrm{e}}

\newcommand{\coupling}{g}
\newcommand{\couplingtd}{\coupling \left( t \right)}
\newcommand{\pulsedur}{\tau}
\newcommand{\ntherm}{n_\mathrm{th}}
\newcommand{\nmech}{n_0}
\newcommand{\ampgain}{\mathfrak{G}}
\newcommand{\ampgainlimit}{\ampgain_{\rm limit}}

\newcommand{\mechdamp}{\gamma}
\newcommand{\mechfreq}{\Omega_{\rm m}}
\newcommand{\detune}{\Delta}
\newcommand{\detunetd}{\Delta \left( t \right)}

\newcommand{\detuneofftd}{\delta \left( t \right)}
\newcommand{\reheatrate}{\Gamma}
\newcommand{\optdamp}{\kappa}
\newcommand{\mineig}{\lambda_{\rm{min}}}
\newcommand{\covmat}{\mathbb{V}}
\newcommand{\gensqz}{S_{\rm{gen}}}
\newcommand{\logneg}{E_{\rm{N}}}

\newcommand{\qangc}{\theta_{\textrm{c}}}
\newcommand{\qangm}{\theta_{\textrm{m}}}
\newcommand{\qangw}{\phi}
\newcommand{\varxgen}{\Var X\s{gen}}

\newcommand{\measfunc}{f^\mathrm{out}}
\newcommand{\measfunctd}{\measfunc \left( t \right)}

\newcommand{\decibel}{\mathrm{dB}}

\lstnewenvironment{python}[1][]
{
    \pythonstyle
    \lstset{#1}
}
{}

\newcommand\pythoninline[1]{{\pythonstyle\lstinline!#1!}}

\newcommand*{\scrpt}[1]{\mathsf{#1}}
\newcommand*{\s}[1]{\ensuremath{_\scrpt{#1}}}
\newcommand*{\up}[1]{\ensuremath{^\scrpt{#1}}}
\newcommand*{\comm}[2]{\ensuremath{ \left[ #1 , #2 \right] }}
\newcommand*{\avg}[1]{\left\langle #1 \right\rangle}
\newcommand*{\sym}[2]{#1 \circ #2}
\newcommand*{\syma}[2]{\avg{\sym{#1}{#2}}}
\newcommand*{\mvec}[1]{\bm{#1}}
\newcommand*{\mmat}[1]{\mathbb{#1}}
\usepackage{mathrsfs} % Also a nice script font package
\newcommand*{\Mode}[1]{\mathscr{#1}}
\newcommand*{\pma}[2]{\begin{pmatrix}#1\\#2\end{pmatrix}}
\newcommand{\rmd}{\mathrm{d}}
\newcommand*{\dd}[1]{\ensuremath{\rmd#1\:}}
\DeclareMathOperator*{\diag}{\operatorname{diag}}
\DeclareMathOperator*{\Var}{\operatorname{Var}}
\usepackage[bbgreekl]{mathbbol} % For blackboard symbols, including lowercase Greek letters
\newcommand{\II}{\mathbb{1}}
\DeclareMathOperator*{\argmax}{arg\,max}

\usepackage[capitalize,poorman]{cleveref}

\begin{document}

\newcommand*{\DeptMathAU}{Department of Mathematics, Aberystwyth University, Penglais Campus, Aberystwyth, SY23 3BZ, Wales, United Kingdom}
\newcommand*{\UPOL}{Department of Optics, Palack{\'y} University, 17.~Listopadu~12, 771~46~Olomouc, Czech~Republic}
\newcommand*{\PhysIC}{Physics Department, Blackett Laboratory, Imperial College London, Prince Consort Road, SW7 2BW, United Kingdom}
\newcommand*{\CEQSMQU}{Center for Engineered Quantum Systems, Dept. of Physics \& Astronomy, Macquarie University, 2109 NSW, Australia}

%original \title{Optimal non-classical correlations of light with a levitated nano-sphere}
\title{Bayesian optimization of non-classical optomechanical correlations}
\author{Alexander \surname{Pitchford}}
\email{agp1@aber.ac.uk}
\affiliation{\PhysIC}
\affiliation{\DeptMathAU}
\affiliation{\CEQSMQU}
%\orcid{0000-0002-4717-2921}

\author{Andrey A. \surname{Rakhubovsky}}
\email{rakhubovsky@optics.upol.cz}
\affiliation{\UPOL}
%\orcid{0000-0001-8643-670X}

\author{Rick \surname{Mukherjee}}
\affiliation{\PhysIC}
%\orcid{0000-0001-9267-4421}

\author{Darren W. \surname{Moore}}
\affiliation{\UPOL}
%\orcid{0000-0002-9228-9619}

\author{Fr\'ed\'eric \surname{Sauvage}}
\affiliation{\PhysIC}
%\orcid{0000-0003-3363-5929}

\author{Daniel \surname{Burgarth}}
\affiliation{\CEQSMQU}
%\orcid{0000-0003-4063-1264}

\author{Radim \surname{Filip}}
\affiliation{\UPOL}
%\orcid{0000-0003-4114-6068}

\author{Florian \surname{Mintert}}
\affiliation{\PhysIC}
%\orcid{0000-0001-8213-4368}

\begin{abstract}
Nonclassical correlations provide a resource for many applications in quantum technology
as well as providing strong evidence that a system is indeed operating in the quantum regime.
Optomechanical systems can be arranged to generate nonclassical correlations (such as quantum entanglement) between the mechanical mode and a mode of travelling light.
Here we propose automated optimization of the production of quantum correlations in such a system,
beyond what can be achieved through analytical methods, by applying Bayesian optimization to the control parameters.
A two-mode optomechanical squeezing experiment is simulated using a detailed theoretical model of the system and the measurable outputs fed to the Bayesian optimization process.
This then modifies the controllable parameters in order to maximize the non-classical two-mode squeezing and its detection,
independently of the inner workings of the model.
We focus on a levitated nano-sphere system, but the techniques described are broadly applicable in optomechanical experiments, and also more widely,
especially where no detailed theoretical treatment is available.
We find that in the experimentally relevant thermal regimes,
the ability to vary and optimize a broad array of control parameters provides access to large values of two-mode squeezing
that would otherwise be difficult or intractable to discover via analytical or trial and error methods.
In particular we observe that modulation of the driving frequency around the resonant sideband allows for stronger nonclassical correlations.
We also observe that our optimization approach finds parameters that allow significant squeezing in the high temperature regime.
This extends the range of experimental setups in which non-classical correlations could be generated beyond the region of high quantum cooperativity.\\[1ex]
\noindent{\bf Keywords: optomechanics, quantum control, entanglement production, Bayesian optimization.\/}
\end{abstract}

\date{\today}
\maketitle

\section{Introduction}\label{sec:Intro}
\begin{figure}
    \includegraphics[width=\columnwidth]{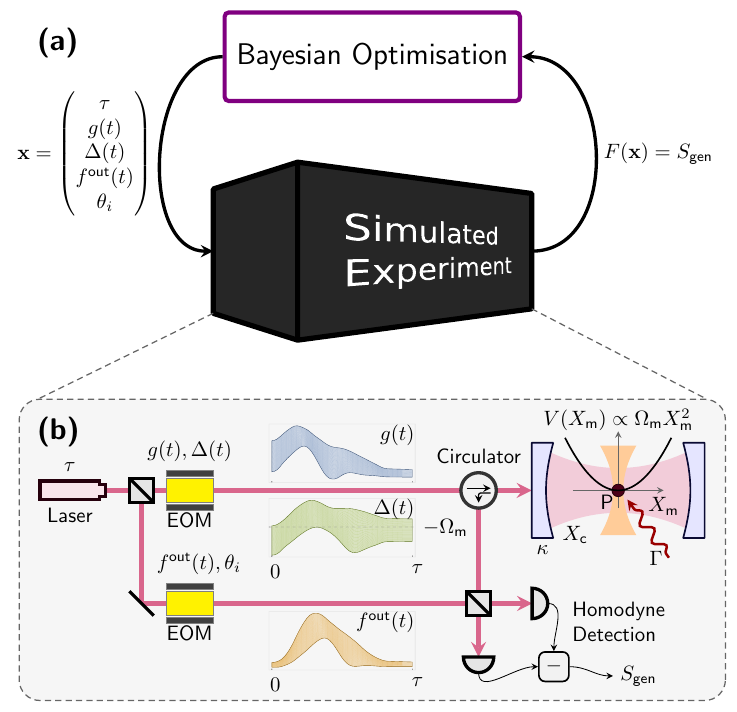}
    \caption{
		(a)~Bayesian optimization is capable of optimizing the squeezing independently of the internal working of the physical
        system, i.e.\ it treats the optomechanics model as a black box, with certain control parameters and an output.
        The control parameters $\mathbf{x}$ (detuning $\detunetd$,
        pulse duration $\pulsedur$ and time-dependent coupling strength $\couplingtd$,
        local oscillator (LO) temporal profile $\measfunctd$) are fed to the black box and produce
        an output figure of merit $F(\mathbf{x})$ (two-mode squeezing).
        Using Bayesian methods, the optimizer searches the parameter space to find the set which maximizes the output,
        thereby discovering the parameters which generate the greatest amount of squeezing.
		\\
		(b)~An example of a levitated optomechanics setup.
        A nanoparticle is suspended via optical tweezers in a harmonic trap.
        The nanoparticle is then surrounded by cavity mirrors and,
        by driving the cavity with an external laser, immersed in a cavity field $X\s{c}$ with linewidth $\kappa$.
        The particle is positioned in the cavity so as to generate a standard optomechanical coupling
        between the mechanical motion $X_{\rm m}$ and the cavity field.
        The pulsed interaction generated by driving on the blue sideband of the cavity
        resonance frequency generates correlations (two-mode squeezing)
        between the cavity field and the mechanical motion.
        The pulse shape and local oscillator profile are controlled by electro-optic modulators (EOM).
        Optimizing over the variables in the external drive provides an enhancement of this effect.
        The output pulse is detected on a homodyne detector (HD) with the local oscillator modulated as per the temporal profile $\measfunctd$.
		The electric signal from the HD is processed classically.
    Repeating the experiment multiple times for a single set $\mathbf{x}_0$ allows gathering statistics sufficient to find the two-mode squeezing $\gensqz = F(\mathbf{x}_0)$.
}
    \label{fig:sketch}
\end{figure}

Nonclassical correlations are necessary to enhance the performance of a variety of quantum technological tasks,
including sensing~\cite{braun2018quantum}, communications and cryptography~\cite{gisin2002quantum},
quantum computing~\cite{Deutsch1985,Jozsa2003,Boyer2017},
quantum thermodynamics~\cite{NonEquilibTherm,PhysRevLett.89.180402,PhysRevX.5.041011},
as well as having significance for foundational questions in quantum physics~\cite{RevModPhys.84.1655}.
Such correlations have been observed in a variety of physical platforms including
optical photons~\cite{chen_experimental_2014,cai_multimode_2017,zhong_12photon_2018,asavanant_generation_2019},
cold atoms~\cite{madjarov_highfidelity_2020,PhysRevLett.112.103601,Omran570},
trapped ions~\cite{Leibfried_interf,Monz,friis_observation_2018},
superconducting circuits~\cite{berkley_entangled_2003,steffen_measurement_2006},
nitrogen-vacancy centers~\cite{dolde_roomtemperature_2013} and,
the platform we address here, optomechanics.

In optomechanical setups, nonclassicality has been observed through the production of
squeezed states of mechanical motion in electromechanical systems~\cite{wollman2015quantum,pirkkalainen2015squeezing},
entanglement between distant mechanical systems coupled by light~\cite{riedinger_remote_2018}
or microwaves~\cite{ockeloen-korppi_stabilized_2018}
and entanglement between the mechanical mode and the microwave mode that leaks from a cavity~\cite{palomaki_entangling_2013}.
This is accomplished by engineering a particular interaction between microwaves and mechanics
through an external classical driving~\cite{hofer_quantum_2011}.
This in turn means that a certain set of experimental conditions
must be satisfied in order for the nonclassical correlations to be generated,
particularly against the deleterious effects of environmental noise.
The determination of these parameters under the constraints of an experimental setting
is a complicated optimization problem even under a small number of tunable variables.

The usual theoretical method to determine the required parameter values is to develop and analyze a mathematical model of the physical system.
Typically, in order to make such a model tractable, many simplifying assumptions must be made.
Further, analytical solutions to the problem are often unavailable and the optimization must proceed numerically.
While a broadly applied technique,
numerical simulation suffers from a structural weakness,
in that the optimization is guided by the accuracy of the mathematical model rather than the experimental data.
Here we invert this viewpoint and propose to use Bayesian methods in optimizing
the production of nonclassical correlations from an optomechanical system.
In our analysis, the optimization variables are the control parameters that drive the actual experiment
and the figure of merit is taken directly from measurements of the system.

\figref{fig:sketch} outlines the process.
The optimization proceeds without any preconceived description of the behavioural response
of the optomechanical setup to any changes in the control parameters $\mathbf{x}$.
This is often referred to as treating the setup as a `black box',
which in this case produces two-mode squeezing (idealized as $F(\mathbf{x})$,
the output of the black box) in response to a set of control parameters $\mathbf{x}$.
By taking advantage of ideas from Bayesian optimization (BO)~\cite{mockus1989bayesian},
the results of the black box itself drive the optimization of the control parameters, not the mathematical model.
This in-place, real-time optimization of the experimental parameters we refer to as
\emph{theory-blind} control optimization.

The theory-blind control idea was pioneered in 1992 under the name `learning control'~\cite{Judson1992}.
Further development and usage has continued, primarily focusing on controlling
chemical reactions~\cite{Phan1997, Phan1999, Weinacht2001, Zhu2003, Cardoza2005, Chen2014}.
There are many implementations and ideas
using a variety of optimization algorithms in the quantum information processing (QIP) field.
An early example used a hybrid approach,
combining classical and theory-blind optimization, to improve gate fidelities~\cite{Egger2014}.
This approach was adopted to improve gate fidelities and reduce drift errors in
single- and two-qubit gates~\cite{Kelly2014, Kelly2016}.
Further applications of theory-blind optimization in QIP can be seen in Refs.~\cite{InSituUpgrade, QVECTOR2017} for example.
The term `model-agnostic' is used widely in the machine learning community with broadly equivalent meaning.
The theory-blind protocol is advantageously applicable in levitated optomechanics experiments,
where a small number of precisely controlled parameters characterize the setup.
This can be used to maximize nonclassical correlations
and is immediately applicable in an experimental setup
such as that used in Ref.~\cite{delic_cooling_2020}.

The possibility to achieve two-mode optomechanical squeezing in a specific levitated optomechanical experiment
was shown in Ref.~\cite{rakhubovsky_detecting_2020},
and further details of the optomechanical theory used in this manuscript can be found there and in the references therein.
Here we take the results of~\cite{rakhubovsky_detecting_2020} as an initial benchmark
and demonstrate that BO is capable of discovering parameter sets that generate significantly stronger two-mode optomechanical squeezing.
This is achieved by efficient exploration of the parameter space,
particularly in the regimes where analytical description of the optomechanical system is difficult or intractable,
specifically, beyond the rotating wave approximation of the resonant-sideband driving,
and outside of the resolved sideband.
Despite the fact that the generation of nonclassical correlations in optomechanics
via a two-mode squeezing interaction is well-investigated theoretically~\cite{genes_robust_2008,hofer_quantum_2011,kiesewetter_scalable_2014,rakhubovsky_robust_2015,Lin2015,rudolph_entangling_2020,lin_entangling_2020},
and has been demonstrated in a number of cryogenic setups~\cite{palomaki_entangling_2013,riedinger_remote_2018},
it remains a challenging task for new optomechanical platforms such as levitated nanoparticles.
Additionally, although some theoretical study has been made into driving correlations through frequencies off the blue sideband~\cite{Lin2015},
only specific fixed frequencies were analyzed.
Other recent work considers driving off the blue sideband~\cite{clarke_generating_2020},
but using very short pulses (less than the period of mechanical motion)
and in the regime where the optical decay is much greater than the frequency of the mechanical oscillator.
Optomechanics with levitated nanoparticles is a promising area of research that in recent years has approached non-classical operation by cooling near the ground state~\cite{delic_cooling_2020,magrini_realtime_2021,tebbenjohanns_quantum_2021,ranfagni_twodimensional_2022,piotrowski_simultaneous_2023,kamba_optical_2022}, and observation of ponderomotive squeezing of light~\cite{magrini_squeezed_2022,militaru_ponderomotive_2022}.
Publications concerning correlations in levitated optomechanics are limited to theoretical proposals~\cite{rudolph_entangling_2020,rudolph_forcegradient_2022,chauhan_tuneable_2022}, and experimental classical correlations~\cite{rieser_tunable_2022,vijayan_cavitymediated_2024}.

Mathematical models of optomechanics are particularly robust and well-tested in the linearized regimes
that our analysis and simulations focus on herein~\cite{aspelmeyer_cavity_2014,bowen_quantum_2015},
thus a successful BO provides predictions on how to maximize
the generation of nonclassical correlations in an optomechanical setting.
Here we allow the BO process access to only the output data indicating nonclassical correlations and the ability to change a set of control parameters of the mathematical model.
The BO therefore cannot tell whether it is working with an experiment or a simulation.
Our study of nonclassical correlations discovered by BO thus constitutes a simulation of the BO applied to a real experiment.
Here, we focus specifically on the parameter regime peculiar to the levitated nanoparticles, however, our methodology can be applied equivalently well to other optomechanical platforms~\cite{patil_measuring_2022a,zivari_onchip_2022,planz_membraneinthemiddle_2023,huang_roomtemperature_2024} given that the latter can operate in the regime of linearized optomechanical coupling and possesses a good sideband resolution (see below).

The remainder of this manuscript is organized as follows.
\secref{sec:Theory} describes the model used in the simulation of the optomechanical system
for calculation of the figure of merit based on the environmental and controllable input parameters.
\secref{sec:Bayes} provides an overview of BO
and explains its suitability for this application.
\secref{sec:OptimResults} gives results of the simulated theory-blind optimization procedure,
demonstrating how increasing the degrees of freedom available to BO 
enables it to discover parameter sets that improve upon the two-mode squeezing levels.
The results and their implications are discussed in~\secref{sec:Discuss}.

\section{Theory}\label{sec:Theory}
In this manuscript we aim at maximizing nonclassical optomechanical correlations.
This section contains a formal description of the optomechanical system formed by a nanoparticle levitated inside a cavity, and a pulse of travelling light.
We provide the Hamiltonian of the optomechanical interaction inside the cavity and obtain the differential equations for the quadratures of the mechanical motion and the intracavity light.
With the help of input-output relations we derive a Lyapunov equation for the matrix of covariances between the mechanical motion and the light pulse, and describe how to quantify the nonclassical correlations between them knowing the covariance matrix.
This section provides the theory necessary to reproduce our results and illustrates which information and control parameters are available to the BO. More details on the theory can be found in Refs.~\cite{genes_quantum_2009,Rakhubovsky2018}. 

\subsection{Gaussian Hamiltonian dynamics of opto-mechanical system}%<<<
\label{sec:OptoMech}
Our focus is on a levitated nanoparticle of mass $\mu\s{p}$ trapped in a tweezer beam within a optical cavity (see~\cref{fig:sketch}~(b)).
In this setup, the potential for the mechanical motion of the particle is determined by the spatial intensity profile of the tweezer.
The Gaussian profile can be well approximated near the origin by a quadratic potential $V (x) = \tfrac 12 \mu\s{p} \mechfreq^2 x^2$ of a harmonic oscillator parametrized by eigenfrequency~$\Omega\s{m}$.
The mechanical motion of the particle is coupled to a cavity mode that itself is a harmonic oscillator characterized by the frequency~$\omega\s{cav}$.

The optomechanical coupling can be introduced in one of two ways depending on the positioning of the nanoparticle inside the cavity and the tweezer polarization.
When the nanoparticle is placed in the antinode of the cavity optical mode, its displacement influences the eigenfrequency of the cavity, which induces the so-called \emph{dispersive} optomechanical coupling~\cite{romero-isart_optically_2011}.
The dispersive optomechanical coupling is inherently nonlinear in the field quadratures~\cite{law_interaction_1995,aspelmeyer_cavity_2014}.
It is, however, typically very weak so in an experiment it is routinely enhanced by a strong coherent driving in the presence of which the interaction is effectively linearized.
Alternatively, when the nanoparticle is in the node of the cavity mode, given an appropriate polarization of the tweezer laser, the optomechanical coupling by \emph{coherent scattering} of the tweezer photons off the nanoparticle into the cavity mode takes place.
Such interaction is linear both in the field and mechanical quadratures~\cite{gonzalez-ballestero_theory_2019}.
It is this type of coupling that allowed ground-state cooling~\cite{delic_cooling_2020} of a levitated nanoparticle.
In both cases, the system can be described by the linearized Hamiltonian of the optomechanical interaction~\cite{aspelmeyer_cavity_2014,bowen_quantum_2015}
\begin{equation}
	\label{eq:hamiltonian}
	\frac{ 1 }{ \hbar } H = \frac 14 \detune (t) ( X\s{c}^2 + Y\s{c}^2 ) + \frac 14 \Omega\s{m} ( X\s{m}^2 + Y\s{m}^2 ) - g (t) X\s{c} X\s{m},
\end{equation}
where $X\s{c}, Y\s{c} ~ (X\s{m}, Y\s{m})$ are the canonical dimensionless quadratures of the cavity (mechanical) mode normalized such that $\comm{X\s{c}}{Y\s{c}} = \comm{X\s{m}}{Y\s{m}} = 2\iu$, and $\detune (t) = \omega\s{cav} - \omega\s{drive} (t)$ is the time-dependent detuning of the coherent drive (or the tweezer frequency for the coherent-scattering coupling) from the cavity frequency.
The coupling strength~$g(t) $ can be set by the power of the coherent drive (or by power and polarization of the tweezer).
In an experiment, the detuning $\Delta (t)$ and the drive power (and consequently, $g(t)$) can be controlled by a suitable modulation (e.g., electro-optical) of the laser light (symbol EOM in~\cref{fig:sketch} where the case of the dispersive optomechanical coupling is pictured).
As we show below, a careful optimization of these parameters allows achieving stronger optomechanical squeezing compared with the primitive regime of constant-power resonant-sideband driving~\cite{rakhubovsky_detecting_2020}.

In this manuscript we are interested in pulsed driving in the vicinity of the upper mechanical sideband of the cavity at frequency $\omega\s{drive} (t) \approx \omega\s{cav} + \Omega\s{m}$.
As is known~\cite{genes_quantum_2009,aspelmeyer_cavity_2014}, driving on the upper mechanical sideband produces an optomechanical interaction which approaches the parametric amplification capable of producing nonclassical correlations by scattering the drive photons to the Stokes sideband.
In order to run efficiently, this process requires that the scattering into the anti-Stokes sideband is suppressed, which occurs when the mechanical frequency exceeds the cavity linewidth: $\mechfreq \gg \kappa$.

We assume a pulsed operation, i.e. $g(t)$ to be nonzero for $0 \leq t \leq \tau$ and zero otherwise.
The advantages of the pulsed manipulation stem from working at shorter timescales compared to the steady states of continuous driving.
Since the pulsed operation does not require the system to reach a steady state, it can use coupling strengths that are prohibitively large for the continuous drive.
Indeed, driving the optomechanical cavity on the upper mechanical sideband adds into the dynamics of the mechanical mode a negative damping proportional to the driving strength~\cite{braginsky_investigation_1970}.
This negative damping easily overwhelms the low intrinsic damping of mechanics thus making its dynamics unstable.
In addition, operating at faster timescales helps to decrease the impact of the noisy thermal environment.

The optomechanical system is open, with each of its modes coupled to its corresponding environment.
Whereas the optical environment has low noise, the mechanical one is at a high temperature.
We take this into account in terms of Langevin-Heisenberg equations in the form~\cite{rakhubovsky_detecting_2020}
\begin{equation}
	\label{eq:lang4}
	\dot{ \mvec v } = \mmat A \mvec v + \mvec \nu,
\end{equation}
where $\mvec v = (X\s{c} , Y\s{c} , X\s{m} , Y\s{m} )$ is a vector of unknowns and $\mvec \nu = ( \sqrt{ 2 \kappa } X\up{in}, \sqrt{ 2 \kappa } Y\up{in} ,  0 , \sqrt{ 2 \gamma } \xi\up{th})$ is a vector of input noises.
In this notation, $\kappa$ is the cavity linewidth and $\gamma$ is the mechanical damping rate.
Note that, following the conventions of~\cite{genes_quantum_2009}, $\kappa$ is an \emph{amplitude} decay rate, and $\gamma$ is the \emph{energy} decay rate.
The drift matrix reads
\begin{equation}
	\mmat A (t) =
	\begin{pmatrix}
		- \kappa      & \detune (t) & 0          & 0        \\
		- \detune (t) & - \kappa    & 2 g (t)    & 0        \\
		0             & 0           & 0          & \Omega\s{m} \\
		2 g (t)       & 0           & - \Omega\s{m} & - \gamma
	\end{pmatrix} \; .
\end{equation}
The components of the noise vector $\mvec \nu$ satisfy the standard Markovian autocorrelations~\cite{giovannetti_phase-noise_2001}
\begin{gather}
	\syma{ Q\up{in} (t) }{ Q\up{in} (t') } = \sigma\s{v} \delta (t - t'), \text{ for } Q = X,Y,
	\\
	\syma{ \xi\up{th} (t) }{ \xi\up{th} (t') } = \sigma\s{v} (2 n\s{th} + 1) \delta (t - t').
\end{gather}
Here $\sym ab = \tfrac 12 ( ab + ba )$ is the Jordan product, $\sigma\s{v} = 1$ is the shot-noise variance, and $n\s{th}$ is the mean occupation of the thermal environment of the nanoparticle.
The nanoparticle's decoherence originates mostly from collisions with residual gas particles, trapping photon recoil and black-body radiation~\cite{romero-isart_optically_2011,delic_cooling_2020}.
A quantity which parametrizes these processes and can be directly estimated from thermalization measurements~\cite{jain_direct_2016} is the heating rate $\Gamma \equiv \gamma n\s{th}$.

An important characteristic of the input fluctuations is the so-called diffusion matrix $\mmat D,$ defined as $\syma{ \mvec \nu_i (t) }{\mvec \nu_j (t') } = \mmat D_{ij } \delta (t - t')$.
In our case
\begin{equation}
	\mmat D \approx \diag ( 2 \kappa , 2 \kappa , 0 , 4 \Gamma ),
\end{equation}
where for the last element we write $ 2 \gamma ( 2 n\s{th} + 1 ) \approx 4 \gamma n\s{th} = 4 \Gamma$.

\subsection{Input-output formalism} %
\label{sec:input_output_formalism}

Since we are interested in control of the nonclassical correlations between mechanics and the leaking light, that can be directed to another quantum system or detector, we have to obtain an expression for the latter.
We start doing so with the input-output relations for a high-$Q$ cavity~\cite{gardiner_input_1985}
\begin{equation}
	\label{eq:input_output}
	\pma{X\up{out}}{Y\up{out}} = - \pma{X\up{in}}{Y\up{in}} + \sqrt{ 2 \kappa } \pma{X\s{c}}{Y\s{c}}.
\end{equation}
Next, we define a mode of the leaking light that is detected at the output.
This mode is characterized by its temporal profile $f\up{out}(t)$ and is described by quadratures
\begin{equation}
	\label{eq:pulses_definition}
	\pma{ \Mode X\up{out} }{ \Mode Y\up{out} } =
	\int_0^\tau \dd s \pma{ X\up{out} (s) }{ Y\up{out} (s) } f\up{out} (s).
\end{equation}
Because the quadratures satisfy the commutation relation
\begin{equation}
	\comm{\Mode X\up{out}}{ \Mode Y\up{out}} = 2 \iu \int_0^\tau \dd s ( f\up{out} (s) )^2 ,
\end{equation}
the mode profile has to satisfy the normalization condition
\begin{equation}
	\int_0^\tau \dd s ( f\up{out} (s) )^2 = 1
\end{equation}
for $\Mode X\up{out}, \Mode Y\up{out}$ to be canonical variables.
In an experiment, the choice of different mode profiles $f\up{out} (t)$, that is detection of quadratures of the modes with different temporal profiles, can be implemented in the homodyne detection by either using a local oscillator with time-dependent amplitude or by frequently sampling the instantaneous value of quadrature with a constant-amplitude local oscillator and subsequently assembling an integral sum of the form~\eqref{eq:pulses_definition} from samples~\cite{takase_complete_2019}.

The choice of a certain temporal detection profile $f\up{out} (t)$ is a particularly important task in the problem of detecting the quantum correlations~\cite{Rakhubovsky2018}.
A simple intuition can be used in the case when the drift matrix is time-independent.
In this case, an analytical solution of the dynamics exists that allows expression of the instantaneous amplitudes of the leaking field $X\up{out} (t), Y\up{out} (t)$ in terms of the initial values and the input fluctuations.
Such an expression contains a term proportional to $X\s{m} (0)$ with the coefficient $T\s{m} (t)$.
Setting the detection profile equal to this coefficient $f\up{out} (t) = T\s{m} (t)$ gives the temporal mode of light that has maximal contribution of $X\s{m} (0)$.
For more details, see e.g. Ref~\cite{Rakhubovsky2018} and~\cref{sec:filtering_functions,sec:readout_of_the_mechanical_state}.

Having the definitions for the output mode, considering it a function of the upper integration limit, we can extend~\cref{eq:lang4} to include the output mode
\begin{equation}
	\label{eq:lang6}
	\dot{ \mvec u } = \mmat B \mvec u + \mvec \mu,
\end{equation}
where $\mvec u = (X\s{c}, Y\s{c}, X\s{m} , Y\s{m} , \Mode X\up{out} , \Mode Y\up{out})$ is the extended $6-$vector of unknowns and $\mvec \mu$ is the extended vector containing noise terms.
\begin{equation}
	\mvec \mu = ( [ \mvec \nu ]_{1\times 4} , - f\up{out} (t) X\up{in}, - f\up{out} (t) Y\up{in} ).
\end{equation}
The new $6 \times 6$ drift matrix reads
\begin{equation}
	\mmat B(t) =
	\left(
		\begin{array}{c|c}
		\left[ \mmat A (t) \right]_{4\times 4} &
		\begin{array}{c}
			0_{2\times 2}
			\\
			\hline
			0_{2 \times 2 }
		\end{array}
		\\
		\hline
		\begin{array}{c|c}
			\sqrt { 2 \kappa } f\up{out} (t) \II_2 & 0 _{ 2\times 2 }
		\end{array}
		                                       &
		0 _{2 \times 2 }
		\end{array}
		\right)
\end{equation}
and for the $6 \times 6$ diffusion matrix we obtain
\begin{multline}
	\mmat F(t)
                                                        \\
	=
	\left(
		\begin{array}{c|c}
			\left[ \mmat D \right]_{4 \times 4}          &
			\begin{array}{c}
				- f\up{out} (t) \sqrt{ 2 \kappa } \II_2
                                                        \\
				\hline
				0_{2 \times 2}
			\end{array}
                                                        \\
			\hline
			\begin{array}{c|c}
				- f\up{out} (t) \sqrt{ 2 \kappa } \II_2  & 0_{2 \times 2 }
			\end{array}
                                                         & f\up{out} (t)^2 \II_2
		\end{array}
	\right).
\end{multline}
Above we used notation $\II_n$ for an $n$-dimensional identity matrix, and $0_{m\times n}$ for a matrix of zeros of corresponding dimensions.

The dynamics of the system are linear, therefore the initial multimode zero-mean Gaussian state and multimode zero-mean Gaussian state of the noises are mapped by~\cref{eq:lang4,eq:lang6} onto another zero-mean Gaussian state.
An important feature of such states is that they are fully described by their second moments that form a covariance matrix.
The latter is defined as
\begin{equation}
	\mmat U_{ij } = \syma{\mvec u_i}{\mvec u_j }. % - \avg{ \mvec u_i } \avg{ \mvec u_j }.
\end{equation}

The covariance matrix $\mmat U (t)$ evolution is governed by the matrix Lyapunov equation:
\begin{equation}\label{eq:covmat_lyapunov}
	\dot{ \mmat U } = \mmat B \mmat U + \mmat U \mmat B\up{T} + \mmat F.
\end{equation}

To analyze the nonclassical optomechanical correlations of the modes of our interest, we derive the covariance matrix of a bipartite system formed by the nanoparticle and the leaking light by keeping only the corresponding rows and columns in $\mmat U$.
In our particular case, we remove the first two rows and columns, and arrive to a $4 \times 4$ covariance matrix $\mmat V_{ij } = \mmat U_{i + 2, j + 2}$ with $1 \leq i,j \leq 4$.

\subsection{Optomechanical two-mode squeezing}
\label{sec:optomechanical_two_mode_squeezing}

From the covariance matrix $\covmat$ of the bipartite optomechanical system we obtain the two-mode squeezing from its minimal eigenvalue $\mineig$.
A squeezed state is indicated by $\mineig < \sigma\s{v} = 1$, with squeezing increasing as $\mineig$ decreases. The two-mode optomechanical squeezing is given by
\begin{equation}\label{eq:GenSqz}
	S_{\rm{gen}} = \max \left\{ 0, -10 \log_{10} \mineig \right\} \, ,
\end{equation}
which is clearly maximized for minimal $\mineig$.

Detection of the two-mode squeezing of a bipartite system does not require full state tomography.
A simple method exists that allows this detection via only one homodyne measurement of each of the two modes.
The method is based on the fact that in the eigenbasis where the covariance matrix is diagonal, the smallest eigenvalue of the covariance matrix is one of its elements.
This means that by a two-mode passive linear transformation it is possible to obtain a generalized quadrature $X\s{gen}$ whose variance equals the smallest eigenvalue of the original covariance matrix~\cite{simon_quantumnoise_1994}.
% This method can exhibit squeezing even when conditional variances do not~\cite{filip_squeezing_2010}.
The most general of such transformations maps the initial quadratures onto a new set, of which we are interested in the one given by
\begin{equation}
	\label{eq:gentwomode}
	X\s{gen} [ \theta\s{c} , \theta\s{m} , \phi ] = X^{\theta\s{c}}\s{c} \cos \phi + X^{\theta\s{m} }\s{m} \sin \phi,
\end{equation}
with $X^{\theta_{i } }_i$ being the quadratures of each subsystem in a rotated basis
\begin{equation}
	X^{\theta}_i = X_{i } \cos \theta + Y_{i } \sin \theta.
\end{equation}
\Cref{eq:gentwomode} thus describes an output quadrature of a virtual beamsplitter having an amplitude transmittance $\cos \phi$ with the rotated quadratures of the original modes as the two input modes.
The variance of $X\s{gen}$ can be computed as
\newcommand*{\vrot}{\overline{ \mmat V }}
\newcommand*{\rr}{\mmat R}
\begin{equation}
    \label{eq:varxgen_from_covmat}
	\Var X\s{gen} = \vrot_{11} \cos^2 \phi + \vrot_{33} \sin^2 \phi + \vrot_{13} \sin 2 \phi,
\end{equation}
where
\begin{equation}
	\vrot = \rr ( \theta\s{c} , \theta\s{m} ) \mmat V \rr ( - \theta\s{c} , - \theta\s{m} ),
\end{equation}
and $\rr$ is the rotation matrix:
\begin{equation}
	\rr( \theta\s{c} , \theta\s{m} ) = \rr_2 (\theta\s{c} ) \oplus \rr_2 (\theta\s{m} ),
	\quad
	\rr_2 (\theta) =
	\begin{pmatrix}
		\cos \theta & \sin \theta
		\\
		- \sin \theta & \cos \theta
	\end{pmatrix}.
\end{equation}
For an optimal set of angles $\theta\s{c}^{(o)} , \theta\s{m} ^{(o)} , \phi^{(o)}$ the corresponding variance assumes the value of the minimal eigenvalue of $\mmat V$:
\begin{equation}
	\Var X\s{gen} [ \theta\s{c}^{(o)} , \theta\s{m}^{(o)} , \phi^{(o)} ] = \lambda\s{min},
\end{equation}

In the lab, one can directly measure $X^\theta_i$ using homodyne detection.
The phase $\theta$ is set by the local oscillator.
The weighting factors $\sin\phi, \cos \phi$ can be optimized offline.
The problem of detecting the two-mode squeezing is then reduced from the full Gaussian tomography of the bipartite state to the \emph{direct} homodyne detection of a pair of quadratures~\cite{rakhubovsky_detecting_2020}.

Note that we simplify the problem of evaluation of the two-mode squeezing by assuming that we have an access directly to the mechanical part of the covariance matrix.
Though technically such a direct access is impossible, the mechanical quadratures can be effectively swapped to a subsequent pulse of leaking light by driving the optomechanical cavity on the lower mechanical sideband $\omega\s{drive} = \omega\s{cav} - \mechfreq$.
The state swap procedure via a red-detuned drive is known to be equivalent to an almost noiseless beamsplitter-like transformation from mechanics to the light~\cite{vanner_towards_2015,rakhubovsky_photon-phonon-photon_2017}.
The problem of the optimal pulse shape is not relevant to the task of state swap as it is for the generation of squeezing.
Therefore, an extension of the problem to include the verification step would only be a technical addition to the problem and would not necessarily extend the scope of the manuscript.
Therefore, we analyze here an optimized upper bound on directly detectable squeezing from the experimental setup with the key time-variable parameters $g(t), \detune (t), f\up{out}(t)$ and~$\tau$.

\section{Bayesian optimization}\label{sec:Bayes}
With the elements of theory developed in the previous section,
both the task of creating and detecting squeezed states in optomechanical systems
can be turned into an optimization problem.
As ultimately these optimizations should be performed directly onto an experimental apparatus,
it is desirable that the optimization routine should converge in a small number of steps,
and exhibit robustness with regards to experimental noise.
Since Bayesian optimization (BO) has been successful with these requirements,
with examples in quantum optimal control
problems~\cite{wigley2016fast, zhu2018training, Henson2018, Nakamura2019, mukherjee_preparation_2020,sauvage_optimal_2020},
it is deemed appropriate for the tasks at hand.

A typical optimization problem involves maximizing a
figure of merit $F(\mathbf{x})$ with respect to control parameters~$\mathbf{x}$,
\begin{equation} \label{optim}
\mathbf{x}^{\rm opt} = \argmax_{\mathbf{x}}~F(\mathbf{x}) .
\end{equation}
In general, $\mathbf{x}$ can be an $N$-dimensional vector where $N$ is the total number of parameters.
In our case, it can describe the control parameters $g(t)$ and $\Delta(t)$ entering the Hamiltonian in~\eqref{eq:hamiltonian},
the pulse duration~$\tau$,
the detection output~$f\up{out}(t)$ profile appearing in~\cref{eq:pulses_definition}
	or the detection angles $\theta$ in~\eqref{eq:gentwomode},
and the figure of merit to be maximized is the two-mode squeezing value in~\eqref{eq:GenSqz}.

This search for optimal control parameters is performed iteratively.
At each iteration of the algorithm the figure of merit %(or its gradients)
is evaluated for a given set of parameters,
from either numerical simulation or experimental data,
and the optimization routine suggests a new set of parameters to be tried.

BO constructs an internal approximation (\emph{model}) of the relationship between the control parameters and the figure of merit $F$,
which guides the optimization process.
The choice of the next set of control parameters is a Bayesian decision problem,
incorporating an incentive to explore the parameter space.
These two steps,
of updating the model based on the full set of evaluations collected and choosing the next set of parameters,
form a single iteration of BO,
and are described briefly below.
More thorough descriptions of BO can be found in
Refs.~\cite{brochu2010tutorial, snoek2012practical, frazier2018tutorial, shahriari2015taking}.

As the optimization of a real experiment is based on a limited number of evaluations,
themselves subject to experimental noise
such as infidelity in applying the controls,
it is convenient to adopt a probabilistic modelling approach wherein
random functions $f$ are used to model the unknown $F$.
The prior distribution $p(f)$ over these random functions is chosen
such that it favors well-defined and regular functions.
Typically this means that $f$ is taken to be a
\emph{Gaussian process} \cite{williams2006gaussian}.
A single evaluation of the figure of merit for control parameters $\mathbf{x}_i$ is denoted $y(\mathbf{x}_i)$
which is allowed to deviate from the true value $F(\mathbf{x}_i)$ due to potential noise in the acquisition of the data.
Then, given a record of $M$ evaluations of the figure of merit
denoted as a vector $\mathbf{y}_M=[y(\mathbf{x}_1),\hdots y(\mathbf{x}_M)]$,
one aims at updating the prior distribution $p(f)$ to take into account the data collected.
This is done by means of Bayes' rule
\begin{equation} \label{update}
p(f|\mathbf{y}_M) = \frac{p(f) p(\mathbf{y}_M|f)}{p(\mathbf{y}_M)} ,
\end{equation}
where the term $p(\mathbf{y}_M|f)$ denotes the likelihood of obtaining the data-set $\mathbf{y}_M$
for a given function $f$ and $p(\mathbf{y}_M)$ acts as a normalization constant.
When the noise in the data is assumed to be normally distributed, and with constant strength,
this conditional distribution can be obtained in closed-form \cite{williams2006gaussian}.

Rather than a single point estimate $f(\mathbf{x})$,
this modelling approach allows to obtain the full probability distribution $p(f(\mathbf{x})|\mathbf{y}_M)$
which can be used to select the next set of parameters $\mathbf{x}_{M+1}$ to be evaluated.
One could choose $\mathbf{x}_{M+1}$ for which the value of $f(\mathbf{x}_{M+1})$ is maximal in average.
However, as the internal BO model is based only on a restricted amount of data, it is likely
that this average value may deviate significantly from the true value of $F$,
especially far away from the parameters already evaluated.
Thus, it is vital to also explore other promising regions of the parameter space.
These considerations can be formulated in terms of an \emph{acquisition function},
which grades a set of pseudo randomly generated potential parameters,
and the choice of the \emph{next} parameters is taken where this acquisition function is maximized \cite{brochu2010tutorial}.

The \emph{Expected Improvement} (EI) acquisition function is the type predominantly used to generate the results presented in \secref{sec:OptimResults},
defined as
\begin{equation}\label{eq:bo_acq_EI}
\alpha_{\rm EI}(\mathbf{x}) = \int_{y_{\rm max}}^\infty \dd y (y-y_{\rm max}) \: p(f(\mathbf{x})=y|\mathbf{y}_M ) \, ,
\end{equation}
where $y_{\rm max}$ is the best evaluation recorded so far.
That is, the evaluation of $f(\mathbf{x})$ that returned the highest figure of merit value.
It effectively quantifies the expected improvement compared with the best recorded output from previous iterations
and encourages exploration where the width of the distribution $p(f(\mathbf{x})|\mathbf{y}_M)$ is large.
For the interested reader, a more detailed explanation of how this EI acquisition function 
achieves this can be found in Ref.~\cite{brochu2010tutorial}
(along with the other acquisition functions used \textit{Lower Confidence Bound} and \textit{Probability of Improvement}).
This exploration feature ensures that a global search is performed, making BO less prone to getting trapped in local minima.

\section{Optimization results}\label{sec:OptimResults}
We look to demonstrate how values for the controllable parameters of the optomechanical system,
specifically those of the driving laser pulse, can be determined automatically using a Bayesian optimization (BO) algorithm.
We aim for parameters that maximize two-mode squeezing using a simulation of the optomechanical setup.
Firstly we show that parameters for a constant (rectangular profile) pulse
that have been shown analytically to maximize two-mode squeezing are also found by the algorithm.
Then we demonstrate that shaping of the coupling strength profile can further increase two-mode squeezing,
and then move on to see that an additional increase is achievable by modulating the detuning frequency.
We also describe simulation results for optimization of homodyne detection measurement angles
and the effects of adding some uncertainty into the strength of the coupling interaction.

The physical parameter values for the simulation are inspired by the setup used to demonstrate the ground state
of an optomechanical system of the same type ~\cite{delic_cooling_2020}.
These parameter values are summarized in~\tblref{tbl:paramRangeTheo}.
The limit for the amplitude gain $\ampgainlimit=50$ is used as a constraint on the optimization variables.
The value is not actually known for the aforementioned experiment,
hence this work is not claiming that the absolute values of generalized squeezing shown are achievable with such a setup,
rather that time-dependent coupling strength and detuning profiles can be found through optimization that improve upon squeezing levels.
Similar results (values for $\gensqz$) are found when simulating with the oscillator initially in thermal equilibrium with the bath, that is $\nmech=\ntherm$.
However, it was found necessary for numerical stability to work with a cooled oscillator at the extreme end of the bath temperature range considered,
so $\nmech=100$ in all simulations unless otherwise specified.

\begin{table}
    \caption{
        \textbf{Optomechanical system parameter values.}
        The default values used for the optomechanical system simulations are the dimensionless values given in the third column.
        Note that the units used here are such that the values are relative to the optical damping, that is where $\optdamp=1$.
        The parameter values are inspired by the experiment~\cite{delic_cooling_2020} where $\optdamp \approx 95$~kHz.
        Equivalent values in standard units for this experiment are given in the last column (note, that
        (i) in~\cite{delic_cooling_2020} $\kappa$ denotes energy damping rate, so $\kappa_\text{\cite{delic_cooling_2020}} = 2 \kappa$  and
        (ii) the trapping frequency is larger which corresponds to better sideband resolution than is assumed in this manuscript).
        The value for the reheating rate $\reheatrate = \mechdamp \ntherm$ is given for the convenience of the reader.}
    \label{tbl:paramRangeTheo}
    \begin{center}
        \small
        \renewcommand{\arraystretch}{1.3}
        \setlength\tabcolsep{5pt}
        \begin{tabular}{ l  l  l  l}
            \thickhline
            symbol          & description                & simulation            & experiment         \\
            \hline
            $\optdamp$      & optical damping            & 1                     & $95$~kHz           \\
            $\mechdamp$     & mechanical damping         & $2.8 \times 10^{-10}$ & $26.6$~Hz          \\
            $\ntherm$       & initial bath phonons       & $2.26 \times 10^8$    & $2.26 \times 10^8$ \\
            $\nmech$        & initial mech. phonons      & $100$                 & $100$              \\
            $\reheatrate$   & reheating rate             & $0.063$               & $6$~kHz            \\
            $\ampgainlimit$ & amplitude gain limit       & $50$                  & $50$               \\
            $\mechfreq$     & mechanical frequency       & $2$                   & $190$~kHz          \\
            $\detune$       & frequency detuning         & $-2$                  & $-190$~kHz         \\ [0.6ex]
            \thickhline
        \end{tabular}
    \end{center}
    
\end{table}

We focus our attention here on maximizing two-mode squeezing.
Similar results can be achieved for maximizing logarithmic negativity,
some of which are given in~\appxref{app:simulation}.
This focus was chosen as there are proposals for more direct measurement of two-mode squeezing in optomechanical experiments.

\subsection{Constant coupling strength}\label{sec:const_coup}

Previous work (analytical, without optimization) has investigated what can be achieved
in terms of two-mode squeezing with a constant (rectangular-shaped) driving pulse~\cite{Rakhubovsky2018,rakhubovsky_detecting_2020}.
The optimization objective is to maximize the figure of merit,
which is the two-mode generalized squeezing $\gensqz$,
with the optimal values of coupling strength $\coupling$ and duration $\pulsedur$ to be determined.
We use the analytically derived profile for the
measurement function $\measfunctd$ described in \secref{sec:input_output_formalism}.
$\gensqz$ is calculated through simulation of the model described in~\secref{sec:Theory}.

The pulse parameters $\coupling$ and $\pulsedur$ are constrained
with experimentally relevant upper bounds of $2\kappa$ and $100 \kappa^{-1}$ respectively.
Moreover, some combinations of pulse parameter values would lead to overheating,
potentially damaging the system,
and so a further constraint $\ampgainlimit=50$ is applied
using the approximation of amplitude gain given by adiabatic regime $\mathfrak G = \me^{ 2 g^2 \tau / \kappa }$
(see~Appendix~\ref{sec:filtering_functions} for more details).

The interactions with the environment are modelled as described in~\secref{sec:Theory}.
The bath temperature is characterized by the mean number of bath phonons $\ntherm$.
Combined with the mechanical damping $\mechdamp=2.8 \times 10^{-10} \optdamp$,
this gives rise to a reheating rate $\reheatrate = \mechdamp \ntherm = 0.063\optdamp$.
The initial temperature of the oscillator is characterized by the mean mechanical occupation $\nmech$.
The model does not predict that cooling of the oscillator significantly increases the
potential two-mode squeezing level $\gensqz$, except for extremely low temperatures ($\ntherm \sim 1$).

\begin{figure}
    \centering
	\begin{tikzpicture}
	\node(a){\includegraphics[width=0.95\linewidth]{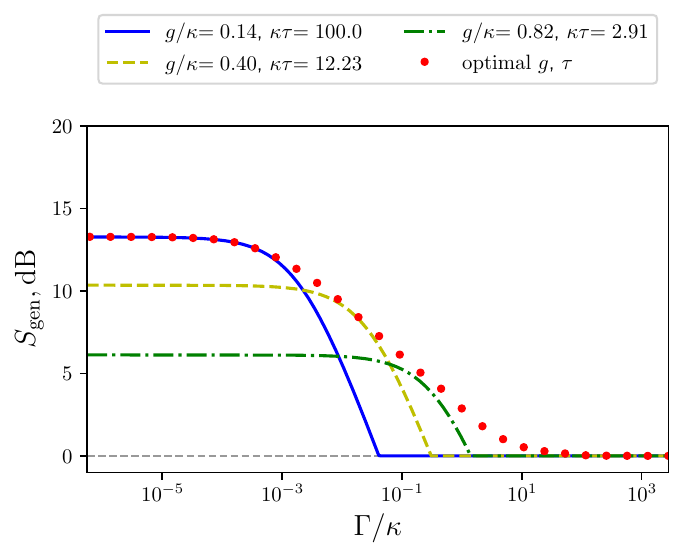}};
	\node at (a.center)
	[
	anchor=south west,
	xshift=4.3mm,
	yshift=-8.5mm
	]
	{
		\includegraphics[width=0.4\linewidth]{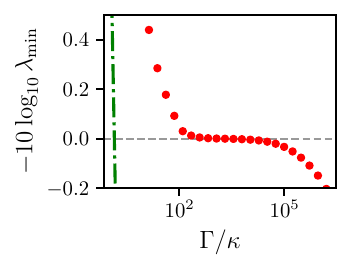}
	};
	\end{tikzpicture}
    \caption{\textbf{Achievable generalized two-mode squeezing over thermal range for constant coupling strength.}
    	The two-mode squeezing $\gensqz$ achieved by optimizing
        constant strength coupling pulse parameters $\coupling$ and $\pulsedur$
        for a given reheating rate $\reheatrate$
        are compared with specific $\coupling,\pulsedur$ values.
        The red dots show the value of the maximum generalized two-mode squeezing
        found through numerical optimization of $\coupling$ and $\pulsedur$.
        The squeezing $\gensqz$ is calculating according to~\eqref{eq:GenSqz}.
        The lines show the squeezing for the fixed specific pulses,
        defined by parameter values shown in the legend.
        The inset is an enlargement and extension of the high heating region of the main plot.
    }
    \label{fig:CompFixedWithOptimTheo}
    % data generated using: srun-ncl-n_ph_range-optim-sko.sh and srun-ncl-n_ph_range-optim-high_heat_rng-sko.sh
    % plotted using: plot_n_ph_sqz.py
    % data in:
    % ../data/n_ph_range-theo_sys-const_optimal-numeric-cool_osc-sgen-sko-try1
    % ../data/n_ph_range-theo_sys-const_optimal-numeric-cool_osc-high_heat_rng-sgen-sko-try1
    % with param files:
    % (main fig)
    % ../parameters/ncl-params-theo_sys-cool_osc-sqz_rng_plt-sko.ini
    % (inset)
	% ../parameters/ncl-params-theo_sys-cool_osc-high_heat_rng-inset_plt-sko.ini
    % data generated with param files:
    % ../parameters/ncl-params-theo_sys-const_optimal-numeric-cool_osc-sgen-sko.ini
    % ../parameters/ncl-params-theo_sys-const_optimal-numeric-cool_osc-high_heat_rng-sgen-sko.ini
\end{figure}

The maximum achievable two-mode squeezing $\gensqz$ is dependent on the reheating rate $\reheatrate$ ~\cite{Rakhubovsky2018}.
This is illustrated in~\figref{fig:CompFixedWithOptimTheo},
which shows the squeezing level achieved by optimizing $\coupling$ and $\pulsedur$
for specific bath temperatures over a wide range.
The relationship of $\gensqz$ to $\reheatrate$ is also plotted for three fixed $\coupling, \pulsedur$ combinations,
with the amplitude gain at maximum $\ampgain = \ampgainlimit$ in each case.
State-of-the-art experiments experience reheating in the range $\reheatrate \approx 10^{-1} \optdamp$.
In this region maximizing two-mode squeezing requires a specific combination of $\coupling$ and $\pulsedur$
that strongly depends on the reheating rate of the mechanical oscillator.

At lower temperatures the squeezing process is adiabatic and hence a long pulse is most effective,
and we find the maximum squeezing $\gensqz$ is determined by the upper bound of the pulse duration $\optdamp \pulsedur \leq 100$.
At the higher temperatures, greater coupling is required to achieve maximal $\gensqz$
and optimal values for coupling strength approach the limit $\coupling / \optdamp \leq 2$.

Reheating rates $\reheatrate \lesssim 10^{-4} \optdamp$ where the squeezing process is practically not influenced by thermal noise
are not achievable in the lab~\cite{delic_cooling_2020,delosriossommer_strong_2021}.
Current experimental setups can achieve reheating rates as low $\reheatrate=0.063 \optdamp$~\cite{delic_cooling_2020},
where we see a strong relationship between $\gensqz$ and $\reheatrate$.
In this reheating region the squeezing level is also most sensitive to pulse parameter values $\coupling$ and $\pulsedur$
(further examined in \appxref{sec:local_traps}).

Due to the sensitivity of two-mode squeezing to the reheating rate
in the experimentally relevant heating range,
if the environmental parameters ($\gamma, n\s{th}$) in an experiment attempting to drive and measure two-mode squeezing
differ from the ones used to theoretically determine the pulse parameters,
maximum possible squeezing will not be achieved.
Theory-blind optimization could be used to determine the correct pulse parameter values
to reach maximal two-mode squeezing levels.

Measurable squeezing is still predicted at higher reheating rates.
For a high bath temperature characterized by phonon number $\ntherm=10^{10}$,
with corresponding reheating rate $\reheatrate = 2.8 \optdamp$,
we find that squeezing $\gensqz = 1.53\,\decibel$ is possible through driving with a constant strength coupling pulse.
With the driving pulse constraints used here, the generalized squeezing becomes effectively immeasurable
($\gensqz < 0.05\,\decibel$)
at some reheating rate limit $\reheatrate \gtrsim 100 \optdamp$.
The squeezing levels plateau in this thermal region, and only actually reach zero for reheating rate $\reheatrate \approx 2000 \optdamp$.
The upper reheating rate limit of measurable squeezing could potentially be raised
if the amplitude gain limit $\ampgainlimit$ could be safely set higher.
Note that this plateauing is related to the mechanical frequency $\mechfreq$,
which is further illustrated and explained in \appxref{sec:hot_bath_num_stab}.

\subsection{Time-dependent coupling}\label{sec:td_coup}
The signal generators used in state-of-the-art optomechanics experiments
allow for effectively any continuous time-dependent function to be applied
to temporally shape the laser pulse amplitude driving the coupling.
That is, with pulse durations on the order of those used here,
it is valid to consider an arbitrarily shaped coupling strength function $\couplingtd$.
Hence the search space for the optimization can be expanded
by adding variables that will allow for a time-dependent coupling strength profile.
The measurement function used for a constant strength driving pulse
cannot be assumed optimal for an arbitrary shaped pulse,
and so any optimization of parameters for the driving pulse
must be combined with optimization of parameters for the measurement function $\measfunctd$.
In a physical experiment such as in~\cref{fig:sketch} these functions, $\couplingtd$ and $\measfunctd$, are controlled by modulators (EOM).

The efficiency of BO, as with any optimization algorithm, is dependent on the number of variables,
so some parameterization scheme must be used to map a small set of variables to a continuous time-dependent function.
We use a piecewise linear (PWL) scheme with 5 equal duration time slices.
Hence the two functions, $\couplingtd$ and $\measfunctd$, are characterized by 6 variables each,
with the total pulse duration $\pulsedur$ as another degree-of-freedom.
The reheating rate $\reheatrate=0.063 \optdamp$ is fixed for the primary comparison of time-dependent coupling
with the constant coupling strength pulse.

\begin{figure}[!ht]
	\centering
	\includegraphics[width=0.99\linewidth]{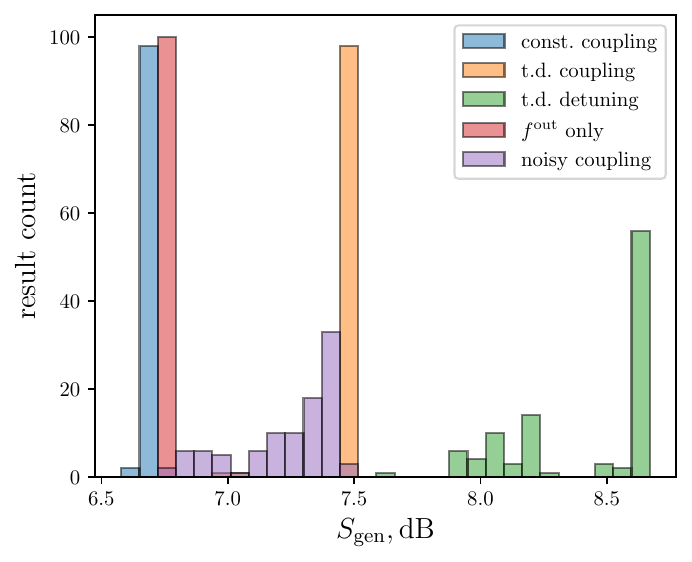}
	\caption{\textbf{Maximized generalized squeezing for coupling pulse variants.}
        The results of maximizing generalized squeezing for the three main different pulse parameterization schemes,
        with increasing degrees of freedom given to the optimization algorithm, are given as:
        `const. coupling' where the strength $\coupling$ and pulse duration $\pulsedur$ are the only variables;
        `t.d. coupling' where the PWL parameters of the time-dependent coupling strength $\couplingtd$ and measurement function $\measfunctd$,
        and the pulse duration $\pulsedur$, are the variables;
        `t.d. detuning' where additionally the PWL parameters of the detuning frequency function $\detunetd$ are also variables.
        Each result set presented in the histograms comprises 100 repeats of the optimization process.
        We see a clear increase in the maximal squeezing as additional degrees of freedom are provided to the algorithm.
        We also see an increase in variation in the outcome as the degrees of freedom increase,
        but in all three schemes greater than 50\% of attempts reach close to the maximal squeezing.
        For the `$\measfunc$ only' set, $\coupling$ and $\pulsedur$ are fixed,
        and only the PWL parameters of the measurement function $\measfunctd$ are optimized.
        For `noisy coupling' optimizations are as `t.d. coupling',
        except that white noise is added to the $\couplingtd$ parameters at each optimization step.
        Interaction with the surrounding environment is characterized by a reheating rate $\reheatrate=0.063 \optdamp$,
        which is within the experimentally relevant range~\cite{delic_cooling_2020}.
	}
	\label{fig:optimal_gen_sqz}
	% plotted using: plot_result_hist.py
	% data:
    % 'const. coupling': ../datatheo_sys-const_optimal-numeric-cool_osc-sgen-sko-hc-optim_results.7443042-combined.txt
    % 't.d. coupling': ../data/theo_sys-5slice_pwl-td_detune-coolish_osc-hotish_bath-optim_results.2318085-combined.txt
    % 'f^{\rm out} only': ../data/theo_sys-5slice_pwl-fout_only-optim_results.185304-combined.txt
    % 't.d. detuning': ../data/theo_sys-5slice_pwl-td_detune-cool_osc-sgen-sko-optim_results.7591232-combined.txt
    % 'noisy coupling': ../data/theo_sys-5slice_pwl-noisy-g0_1-optim_results.182698-combined.txt
	% with param file: ../parameters/ncl-params-theo_sys-sgen-sko.ini
    % data generated with param files:
    % ../parameters/ncl-params-theo_sys-const_optimal-numeric-cool_osc-sgen-sko-hc.ini
    % ../parameters/ncl-params-theo_sys-5slice_pwl-cool_osc-sgen-sko.ini
    % ../parameters/ncl-params-theo_sys-5slice_pwl-td_detune-cool_osc-sgen-sko.ini
    % ../parameters/ncl-params-theo_sys-5slice_pwl-fout_only-sgen-sko.ini
    % ../parameters/ncl-params-theo_sys-5slice_pwl-cool_osc-noisy-sgen-sko.ini
\end{figure}

The optimization results comparing the constant and time-dependent coupling are shown in~\figref{fig:optimal_gen_sqz}.
There is some variation in the value of $\gensqz$ found by BO,
and so the results of 100 repeats of the optimization process are shown in histograms.
%The data from the different optimization variable combinations are compared on the same axes.
%The results for the top-hat and time-dependent coupling pulses
%are labelled `const. coupling' and `t.d. coupling' respectively.
In all repetitions the time-dependent coupling strength pulses out-perform the constant strength pulses,
with maximum squeezing of $\gensqz = 7.51\,\decibel$ for the PWL coupling strength,
compared with $\gensqz = 6.66\,\decibel$ for constant coupling strength,
demonstrating that greater squeezing can be achieved
with a temporally shaped coupling strength.
For the PWL coupling strength, 98\% of repetitions are within $0.06~\decibel$ of the maximum,
indicating that only a few repeats would be necessary to achieve near maximum possible squeezing.

We also find that at higher reheating rates, optimized PWL shaping increases the level of squeezing
beyond what can be achieved with a constant strength coupling pulse.
For example, with reheating rate $\reheatrate = 2.8 \optdamp$,
we find that the PWL parameterization of the coupling pulse allows squeezing up to $\gensqz = 1.81\,\decibel$,
compared with $1.53~\decibel$ for the constant strength.

The data labelled `$\measfunc$ only' show results of optimizations where the
coupling strength $\coupling$ and duration $\pulsedur$ are fixed at the optimal values,
and only the PWL parameters of the measurement function $\measfunctd$ are optimized.
We see a small improvement in squeezing over the analytically derived measurement function,
but clearly the vast majority of the increase in squeezing comes from modulating the coupling strength.
The optimal measurement function for the constant coupling strength pulse was derived using a rotating wave approximation (RWA).
In the simulations here we use a numerical solver and so the RWA is not necessary.
This explains the small improvement we see between the constant coupling and `$\measfunc$ only' results
(with further details given in \appxref{sec:local_traps}).

\subsection{Detuning frequencies}\label{sec:detuning}
The results presented so far have been obtained
for a fixed driving laser frequency detuning $\detune = -\mechfreq$.
Solving the dynamics numerically also allows for the detuning $\detune$
to be offered as a variable for optimization. 
If the optimized detuning is held constant throughout the duration of the driving pulse,
only a small improvement is found in the maximum achievable squeezing,
with $\gensqz = 6.73\,\decibel$ (at $\detune \approx -(1.05)\mechfreq$),
compared with $\gensqz = 6.66\,\decibel$ for fixed blue-sideband detuning of the time-dependent strength pulse.
However, allowing a time-dependent profile for the detuning,
by optimizing the parameters of a PWL function $\detunetd$,
enables significantly greater two-mode squeezing.

The results for repeated optimizations maximizing $\gensqz$ including the PWL detuning variables
can be seen in~\figref{fig:optimal_gen_sqz}, labelled `t.d.~detuning'.
To be clear, the optimization variables in this case are the PWL parameters of the functions
$\couplingtd$, $\measfunctd$, $\detunetd$ and the duration $\pulsedur$.
The optimization of the detuning frequency is constrained to $-\frac32\mechfreq \le \detune \le -\frac12\mechfreq$ for each PWL point.
The maximum achieved squeezing is $8.67~\decibel$,
significantly higher than that found with the detuning fixed at $\detune = -\mechfreq$.
There is greater variability in the optimization result,
but $\gtrsim 50\%$ of the attempts find maximal squeezing within $0.05~\decibel$ of the highest value found.

We find that temporal shaping of the detuning also increases squeezing at higher reheating rates.
For the high bath temperature,
with reheating rate $\reheatrate = 2.8 \optdamp$,
we find that the optimized detuning frequency modulation allows squeezing up to $\gensqz = 2.16\,\decibel$,
compared with $1.81~\decibel$ for fixed blue sideband detuning.
Illustration of this comparison is given in \appxref{sec:hot_bath_num_stab}.
We emphasize that all types of pulse are bound by a limit $\ampgainlimit$ on the amplitude gain,
and so this comparison is made at the physical (overheating) limit of the optomechanical system.

\begin{figure}
    \centering
    \begin{minipage}{\linewidth}
        \centering
        \includegraphics[width=0.9\textwidth]{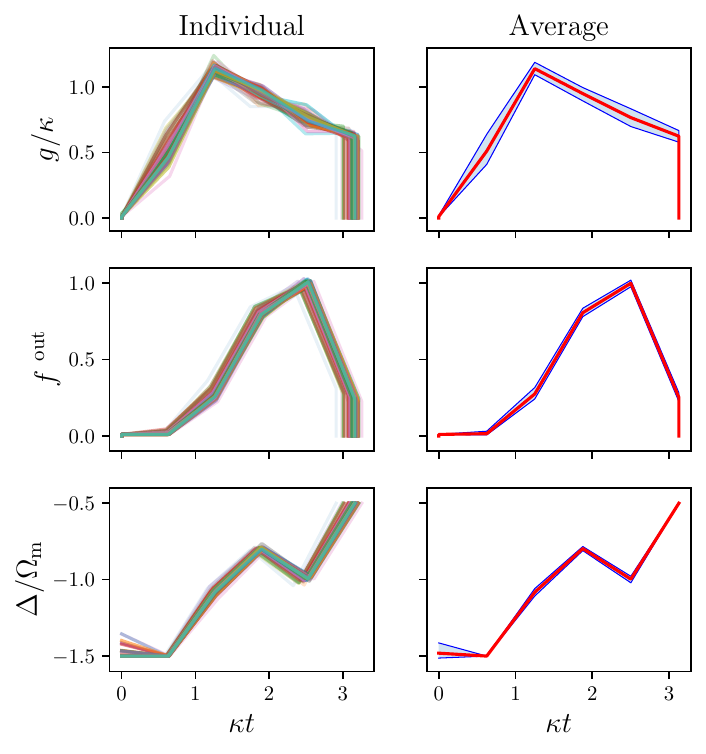}
    \end{minipage}
    \caption{\textbf{Optimized pulses with variable coupling, measurement and detuning.}
        The individual and average optimal function profiles are shown for the best 50 results
        optimizing the parameters of the time-dependent profiles of the controllable variables:
        coupling strength $\couplingtd$,
        measurement function $\measfunctd$
        and laser field frequency detuning $\detunetd$.
        The detuning is shown relative to the mechanical frequency $\mechfreq$,
        such that $\detune / \mechfreq = -1$ corresponds to the resonant (blue sideband) frequency.
        The mean value of the amplitude at the timeslot boundaries,
        and the mean pulse duration,
        are used to plot the average pulse profiles (solid red lines).
        The average profiles are bounded (blue lines) by twice the standard deviation in the pulse parameters.
        The pulses correspond to the optimized generalized squeezing
        dataset labelled `t.d.~detuning' in~\figref{fig:optimal_gen_sqz}.
    }
    \label{fig:detuning_pulses}
    % plotted using: plot_pulse.py
    % data:
    % ../data/theo_sys-5slice_pwl-td_detune-cool_osc-sgen-sko-optim_results.7591232-combined.txt
    % with param files:
    % ../parameters/ncl-params-theo_sys-5slice_pwl-td_detune-cool_osc-sgen-sko.ini
\end{figure}

\figref{fig:detuning_pulses} shows the time-dependent profiles of the optimal
$\couplingtd$, $\measfunctd$ and $\detunetd$.
The plots show representations of the individual pulses that gave rise to the 50 highest squeezing results
and the average pulse temporal profiles for coupling strength, measurement and detuning.
All the profile sets included in the averages produced squeezing within $0.05~\decibel$ of the highest $\gensqz$ value found.
The average pulse itself produces squeezing within $0.01~\decibel$ of the highest value.
The similarity between profiles of the optimized pulses hints at the existence of a uniquely optimal solution,
which may be accessible analytically.

The profiles exhibit distinct features,
with the coupling strength $\couplingtd$ rising from zero to an early peak,
then tailing off linearly to an abrupt finish at half maximum amplitude.
The measurement function $\measfunctd$ rises at first exponentially,
then more gently, to a late peak, dropping to near zero by the end.
The detuning $\detunetd$ starts at its lower bound,
then from the second timeslot, rises to towards its upper bound
(where it ends), but with a dip that corresponds to the peak in $\measfunctd$.
Functions with these features could be used in experimental attempts
without necessarily the need for automated optimization within the experiment.

The particular result for the detuning temporal profile is interesting,
as most analytical studies have assumed fixed detuning at the blue sideband for driving two-mode squeezing.
Possibly, the time-dependent detuning frequency helps counteract the noise,
and hence leads to greater squeezing.
Some evidence for this is observed when setting the mechanical oscillator frequency
much greater than the optical damping ($\mechfreq = 50 \optdamp$).
The optimized time-dependent profile for the detuning $\detunetd$ is
then at the blue sideband when the coupling is at its strongest.

\subsection{Detection angles}
\label{sec:detect}
In an experimental setting, estimating the covariance matrix $\covmat$ would require full state tomography.
There is a potentially more efficient method of measuring $\gensqz$,
described in~\secref{sec:optomechanical_two_mode_squeezing},
based on the equivalence of $\varxgen\left[\qangc, \qangm, \qangw \right]$ to $\mineig$.
This method requires three parameters to be determined: $\qangc, \qangm, \qangw$.
The latter is a weighting that can be determined in post processing.
The homodyne measurement angles $\qangc, \qangm$ are experimental settings
that could be determined through theory-blind type optimization.
This would require repeated driving with the same pulse parameters,
taking measurements to calculate $\varxgen$ as per~\eqref{eq:varxgen_from_covmat},
and using the optimization algorithm to find the values of $\qangc, \qangm$ that minimize $\varxgen$.

Tests in simulation find that a gradient descent algorithm (BFGS) finds the optimal values of $\qangc, \qangm$,
such that $\varxgen\left[\qangc, \qangm, \qangw \right]=\mineig$ to an acceptable level of precision.
With reheating rate $\reheatrate=0.063 \optdamp$
it is necessary to cool the oscillator, for instance such that $\nmech=100$.
Much lower $\nmech$ values than this are achievable in experiment~\cite{delic_cooling_2020}.
Optimization attempts with the oscillator in initial thermal equilibrium $\nmech=\ntherm$
are not found to be reliable in locating the global minimum for $\varxgen$.
Explanation for this, and details of the optimization method, are given in~\appxref{sec:detect_optim}.

\subsection{Control noise}
\label{sec:noise}
In an experimental setting the precision to which controls can be applied will be limited.
Also, how the system will respond to the controls may not be fully predictable.
In this specific example of controlling the coupling of the oscillator
to the light field by modulating the amplitude of the laser,
it is likely that the actual coupling may have some random variation in its response.
This is referred to as \emph{control noise}.
The optimization algorithm is guided by the outcome of trying specific sets of parameters.
Variation in the outcome will lead to reduced performance of the algorithm.
One method to overcome this would be to repeat the experiment
with the same parameters multiple times and take the mean outcome,
however this would greatly increase the total number of times the experiment would need to be run.
Bayesian optimization can refine its model of the control landscape,
taking into account that the figure-of-merit function value for some set of parameters may not be exact.

To replicate this scenario that one would encounter in the lab,
and to verify the stability of optimization,
the control noise is modelled by adding some
pseudo-random Gaussian distributed value to the pulse parameters
(further details are given in~\appxref{sec:noisy_optim}).
The results for maximizing $\gensqz$ by optimizing $\couplingtd$ and $\measfunctd$
when noise (standard deviation $10\%$ of amplitude)
is added to the piecewise linear parameters of $\couplingtd$
are shown in~\figref{fig:optimal_gen_sqz},
labelled `noisy coupling'.
The maximum squeezing achieved with these noisy controls is $\gensqz = 7.47\,\decibel$,
which is within $0.05~\decibel$ of the maximum achieved without control noise.
We see greater variance in the outcome of the optimization,
but still $\gtrsim 50\%$ of attempts within $0.2~\decibel$ of the maximum.

\section{Summary and Outlook}\label{sec:Discuss}
The simulations have demonstrated the effectiveness of automated optimization in the production of nonclassical optomechanical correlations,
as measured by the two-mode squeezing~$\gensqz$.
This is accomplished by adding a layer of Bayesian optimization to the control variables,
such as coupling rate $\couplingtd$, pulsed interaction duration $\pulsedur$, drive detuning $\detunetd$ and the detection profile $\measfunctd$,
so that repeated simulations of the optomechanics experiment are directed towards increased nonclassical correlations.
Such controllable variables are optimized against the uncontrolled parameters such as the heating rate $\reheatrate$
that negatively affect the production of nonclassical correlations.

For example, in the case of a pulse of constant interaction strength
the optimization distinguishes between various reheating regimes and pulse lengths $\pulsedur$ in order to maximize the optomechanical two-mode squeezing~$\gensqz$
in experimentally relevant regions of the reheating variable.
Adding more variables to the optimization procedure, including time-dependent coupling strength~$\couplingtd$, detuning~$\detunetd$ and measurement functions~$\measfunctd$,
only increases the effectiveness of the optimization procedure.
While requiring more resources to optimize,
having a wider parameter landscape to explore provides more opportunity to increase the optomechanical squeezing.
We have assumed a certain fixed complexity of this time-dependence in the form of piecewise linear functions,
however it seems reasonable to conjecture that increasing the detail of such functions,
and therefore the number of control variables,
will produce more finely tuned optimizations with greater nonclassical correlations.

We find that time-dependent detuning away from the blue-sideband,
in combination with the other optimized variables,
produces noticeably greater squeezing than otherwise predicted~\cite{rakhubovsky_detecting_2020}.
The blue-detuned drive produces nonclassical correlations perfectly in a unitary system,
and we deviate from this by including noise effects from the thermal environment.
Allowing the control variables to vary around this unitary ideal
gives the optimization an opportunity to locate the deviated maximum squeezing.
For an estimate, while the non-optimized case predicts generation of approximately $6.7$~dB 
of the optomechanical squeezing for the heating rate $\Gamma = 0.063 \kappa$, 
after the optimization over all experimentally controllable parameters 
($\pulsedur, \couplingtd, \measfunctd, \detunetd$) the squeezing can reach values over $8.6$~dB.

The optimization not only increases the magnitude of the optomechanical squeezing but,
compared to the non-optimized case~\cite{rakhubovsky_detecting_2020},
allows for greater squeezing at all relevant environment temperatures up to the measurable squeezing threshold ($\gensqz \gtrsim 0.05\,\decibel$).
However, we do not find that the upper limit of the reheating rate $\Gamma$ 
(that allows generation of measurable optomechanical squeezing)
is raised by time-dependent modulation of the driving parameters.
In considering higher temperature environments, we have focused on the maximization of squeezing at a fixed, experimentally reasonable,
value of the temperature of the environment corresponding to a certain value of $\Gamma = 2.8 \kappa$.
At this high temperature of the mechanical environment,
driving the system with a constant coupling strength yields $1.53$~dB of squeezing.
For the same reheating rate, the pulse with all the parameters ($\pulsedur, \couplingtd, \measfunctd, \detunetd$)
optimized can exhibit detectable squeezing as high as $2.16\,\decibel$.
Note that from the point of view of the quantum cooperativity~\cite{aspelmeyer_cavity_2014},
which compares the optomechanical coupling rate with the decoherence rates of the system $C_q = 2 g^2 / ( \kappa \Gamma) \lesssim 0.8 $,
at this high temperature, our system is in the regime of moderate cooperativity.
The key property here is the relationship of the coupling and heating rates: $g / \Gamma \lesssim 20$, much lower compared to prior experimental reports of opto-/electromechanical continuous-variable entanglement.
For example, in~\cite{ockeloen-korppi_stabilized_2018},~$g / \Gamma \lesssim 73$, in~\cite{riedinger_nonclassical_2016},~$g / \Gamma \lesssim 2\times10^3$.

% Reference parameter values:
% Ockeloen-Korppi
% \gamma = 2 \pi 100 Hz; nth = 41, g = (up to ) 2 pi 300 kHz 
% => g / \Gamma = 73
% Riedinger
% g0 = 2 pi 825 kHz; omega_m = 2 pi 5.3 GHz , Q_m = 1.1 10^6, gamma_m = omega_m / Q = 4.82 kHz;  nth = 0.1
% => g / \Gamma = 1711.618257

While we have emphasized experimental values from a certain levitated
setup~\cite{delic_cooling_2020} in the text,
the present simulations can be applied to other levitated
experiments~\cite{monteiro_dynamics_2013,hempston_force_2017,pontin_levitated_2018,meyer_resolvedsideband_2019}.
Moreover, any optomechanical~\cite{meenehan_silicon_2014,nielsen_multimode_2017,shomroni_twotone_2019}
or electromechanical~\cite{barzanjeh_stationary_2019,peterson_ultrastrong_2019}
system capable of functioning in the pulsed regime of linearized optomechanical interaction
can be analysed using exactly the same tools.
The latter also allow consideration of multiple mechanical oscillators
for investigation of mechanical-mechanical correlations generated by the pulsed interaction.
Furthermore, the problem can be extended to include two- and three-dimensional motion of the levitated nanoparticles,
which allows consideration of non-classical correlations
between the motion of one or multiple nanoparticles in orthogonal directions~\cite{vanloock2003detecting,chang2020observation}.

Bayesian optimisation (BO) has been found to be very effective in locating optimal parameters for maximizing two-mode squeezing
and logarithmic negativity, which are both recognized measures of nonclassicality in optomechanical systems.
The optimization landscape was found to contain many sub-optimal local maxima (local traps).
BO is highly effective at navigating around these traps to find a solution that is globally optimal.
For higher dimensional landscapes, 
such as those needed for finding solutions to the full time-dependent parametrization of coupling strength, measurement function and detuning,
BO requires increasingly more figure-of-merit evaluations, which equate to experimental runs in an automated optimization setting.
However, the number of evaluations is vastly reduced over using a gradient-based algorithm on a densely-trapped optimization landscape.
There is scope for investigating algorithms that learn general characteristics of the optimization landscape to improve efficiency in specific searches.
Success with algorithms of this type in similar applications have been reported~\cite{Moon2020,Dalgaard2020}.
This approach could further reduce the number of figure-of-merit evaluations,
which is especially advantageous when directly optimizing controllable parameters of a physical system,
but is expected to be more computationally intensive than BO.

The advantages of this theory-blind optimization are threefold: 1) We see improvement in the optimization of the mathematical model of an optomechanical system even though such models are very well explored. As mentioned, in the simulation we reliably find near identical profiles for $\couplingtd, \measfunctd, \detunetd$ in each optimization attempt and this may guide theoretical work in searching for an analytical basis for our optimized values. 2) We have, in principle, an improvement in experimental control that goes beyond the theoretical intuition granted by mathematical models and can be fully automated. While our main point is that BO can be applied blindly to the experimental setup to achieve such improvements, we also note that due to the consistency of results from point 1), the optimized functions could be directly attempted experimentally to verify the expected high squeezing without the interfacing of the BO with the experiment. Finally, 3) the theory blind nature of our results suggests that experimental improvements can be achieved without any theoretical intuition or known mathematical model. In this case, the attainment of a given figure of merit may in fact navigate mathematical modelling. In particular, we aim point 3) at systems in which nonlinearity plays a strong role, such that intuition based on linear systems is poor and mathematical models are typically intractable. A caveat here is that we have only demonstrated the effectiveness of BO for a linear system. However in recent years the explosion of machine learning methods has also been applied to nonlinear quantum systems~\cite{weiss_quantum_2019}, showing that such objections may indeed be overcome.

\section{Acknowledgements}
A.R., D.M., and R.F. acknowledge discussions with M. Aspelmeyer, N. Kiesel and U. Deli{\' c}.
The Supercomputing Wales clusters were used to run the many simulations behind the results.
We are grateful for the free access to this resource.
The project Theory-Blind Quantum Control {\it TheBlinQC} has received funding from
the QuantERA ERA-NET Cofund in Quantum Technologies implemented
within the European Union's Horizon 2020 Programme and from EPSRC under the grant EP/R044082/1.\\
A.A.R, D.M. and R.F. were supported by the Czech Scientific Foundation (project 23-06308S),
the MEYS of the Czech Republic (grant agreement No 731473),
the Czech Ministry of Education
(project LTC17086 of INTER-EXCELLENCE program)
and have received national funding from the MEYS
and the funding from European Union’s Horizon 2020 (2014-2020)
research and innovation framework programme under grant agreement No. 731473
(project 8C18003 TheBlinQC).
F.S. was supported through a studentship in the  Quantum  Systems  Engineering Skills and Training Hub at Imperial College London funded by EPSRC(EP/P510257/1).

\appendix

\section{Nonclassicality measures}
\label{app:measures}
In this appendix we elaborate on our choice of the figure of merit for nonclassical correlations in optomechanics.

First, we prove that positive two-mode squeezing (TMS) is a witness of entanglement given that neither of the individual modes shows single-mode squeezing.
Separable states are such that admit representation of their density matrix in the form of a mixture of product states~\cite{plenio_logarithmic_2005}:
\begin{equation}
  \label{eq:define_separable}
  \rho\s{sep} = \sum_i p_i \rho^{(i)}\s c \otimes \rho^{(i)}\s m.
\end{equation}
States that are not separable (do not admit such representation) are called \emph{entangled}.
The variance of a generalized quadrature~\eqref{eq:gentwomode}, evaluated a separable state of the form~\eqref{eq:define_separable}, reads (we omit superscripts $X\s m ^{\theta_c} \to X\s m$ for simplicity of notation)
\begin{multline}
  \avg{ \Delta ( X\s{gen} )^2 } = \avg{ X\s{gen} ^2 } - \avg{ X\s{gen} }^2 =
  \\
  \avg{ \cos^2 \phi X\s{c}^2 + \cos \phi \sin \phi ( X\s{c} X\s{m} + X\s{m} X\s{c} ) + \sin^2 X\s{m}^2 }
  -
  \\
  \cos^2 \phi \avg{ X\s{c}}^2 - 2 \cos \phi \sin \phi \avg{ X\s{c} }\avg{ X\s m } - \sin^2 \phi \avg{ X\s{m}}^2
  =
  \\
  \cos^2 \phi \avg{ \Delta X\s{c} ^2 } + \sin^2 \phi \avg{ \Delta X\s m ^2 }.
\end{multline}
In the last line we used the fact that for the product states and mixtures thereof, $\avg{ X\s m X\s c } = \avg{ X\s m } \avg{ X\s c }$.

Lack of squeezing in individual modes means that for both subsystems, $\avg{ \Delta X_i ^2 } \geq 1$.
Consequently,
\begin{equation}
  \avg{ \Delta X\s{gen}^2 } \geq \cos^2 \phi + \sin^2 \phi = 1.
\end{equation}
That is, for states that admit expansion like~\eqref{eq:define_separable}, the variance of a generalized quadrature, and consequently, the smallest eigenvalue of the covariance matrix, is bounded below by the shot-noise variance (given that individual modes are not single-mode squeezed).

A necessary and sufficient criterion for observation of entanglement can be derived using logarithmic negativity~\cite{plenio_logarithmic_2005}.
The logarithmic negativity (LN) is an entanglement monotone which for Gaussian states admits a convenient representation in terms of the elements of the covariance matrix.
A Gaussian state of an optomechanical system can be described by a covariance matrix $\mmat V$ written in the block form
\begin{equation}
  \mmat V =
  \begin{pmatrix}
    V\s m & C
    \\
    C^\top & V\s c
  \end{pmatrix},
\end{equation}
where $V\s{m}, \ V\s c, \ C$ have dimensions $2 \times 2 $ each, and contain, respectively, variances of the mechanical mode, optical mode and cross-correlations between the modes.
Given such covariance matrix, the LN can be expressed as~\cite{laurat_entanglement_2005}
\begin{equation}
  E\s N = \max \left[   0 , \dfrac 12 \log_2 \dfrac 12 \left( \Sigma_{\mmat V} - \sqrt{ \Sigma_{\mmat V}^2 - 4 \det \mmat V } \right) \right],
\end{equation}
where
\begin{equation}
  \Sigma_{\mmat V} = \det V\s m + \det V\s c - 2 \det C.
\end{equation}

The advantage of using LN is that it occurs to be a necessary and sufficient criterion of entanglement.
On the practical side, estimation of the LN requires knowledge of the full covariance matrix of the bipartite state which, in general case, has 10 independent elements.
Estimation of the TMS requires knowledge of only three of these elements.
This suggests that, given the knowledge of the model that allows to choose the elements to estimate wisely, it is possible to decrease the experimental cost of the entanglement verification.

\section{Approximations in the theory of cavity optomechanics}
\label{app:approximations}
\subsection{Dynamics of slow opto-mechanical amplitudes}
\label{sec:dynamics_of_slow_amplitudes}

To investigate the interaction between the optical and mechanical modes, it is instructive to switch to the rotating frame defined by the free evolution of the modes (the first two terms in~\cref{eq:hamiltonian}).
This is equivalent to a transition from the instantaneous values of the quadratures $\mvec v$ to their slowly varying envelopes $\mvec v\up{env}$ following the rule
\begin{equation}
	\label{eq:slowamplitudes}
	\mvec v = \mmat R \mvec v\up{env}
	\end{equation}
with $\mmat R = \mmat R_2 ( \detune t ) \oplus \mmat R_2 (\mechfreq t)$, where $\mmat R_2$ is the matrix of unitary rotation
\begin{equation}
	\label{eq:small_rotation}
	\mmat R_2 ( \alpha) =
	\begin{pmatrix}
		\cos \alpha & \sin \alpha \\
		- \sin \alpha & \cos \alpha
	\end{pmatrix}.
\end{equation}

Substituting~\cref{eq:slowamplitudes} into~\cref{eq:lang4} yields the equations of motion for the envelopes $\mvec v\up{env}$:
\begin{equation}
	\label{eq:lang_env}
	\dot{ \mvec v }\up{env} = \mmat A\up{env} \mvec v\up{env} + \mvec \nu \up{env},
\end{equation}
completely analogous to~\cref{eq:lang4} with notations
\begin{equation}
	\mmat A\up{env} = \mmat R^{-1} ( \mmat A \mmat R - \dot{ \mmat R } ),
	\quad
	\mvec \nu \up{env} = \mmat R^{-1} \mvec \nu.
\end{equation}
In particular, the full expression for $\mmat A\up{env}$ reads
\begin{equation}
	\mmat A\up{env} =
	\begin{pmatrix}
		- \kappa \cdot \II_2  & g \cdot \mmat \Theta ( \detune , \mechfreq )
		\\
		g \cdot \mmat \Theta ( \mechfreq , \detune ) &  - \tfrac \gamma 2 \cdot \left( \II_2 + \mmat \Phi (\mechfreq) \right)
	\end{pmatrix},
\end{equation}
with notation
\begin{gather}
	\mathbb \Theta (\detune , \mechfreq ) =
	\begin{pmatrix}
		- 2 \cos ( \mechfreq t )  \sin ( \detune t)	, - 2 \sin ( \mechfreq t ) \sin ( \detune t )
		\\
		2 \cos ( \mechfreq t ) \cos ( \detune t ) , 2 \sin ( \mechfreq t ) \cos ( \detune t )
	\end{pmatrix},
	\\
	\mathbb \Phi (\mechfreq) =
	\begin{pmatrix}
		- \cos (2 \mechfreq t ) & - \sin ( 2 \mechfreq t )
		\\
		- \sin ( 2 \mechfreq t ) & \cos ( 2 \mechfreq t )
	\end{pmatrix}.
\end{gather}
Also for the noises one can write
\begin{gather}
	\mmat D\up{env} =
	\begin{pmatrix}
		2 \kappa \cdot \II_2 & 0_{2\times 2}
		\\
        0_{2\times 2} & 2 \Gamma \cdot ( \II_2 +  \mathbb \Psi)
	\end{pmatrix},
    \\
	\mathbb \Psi =
	\begin{pmatrix}
		- \cos 2 \mechfreq t & - \sin 2 \mechfreq t
		\\
		- \sin 2 \mechfreq t & \cos 2 \mechfreq t
	\end{pmatrix}
\end{gather}
A Lyapunov equation with the matrices substituted according to the rule $ \bullet \mapsto \bullet \up{env}$ can be written for the equations of motion~\cref{eq:lang_env}.
It is important to note that this equation is exact and valid for an arbitrary detuning.

\subsection{Rotating wave approximation}
\label{sec:rotating_wave_approximation}

To simplify the further analysis, we assume that the system is operated in the resolved-sideband regime, where the mechanical frequency significantly exceeds the linewidth of the cavity, and the opto-mechanical coupling is weak: $\mechfreq \gg \optdamp , \coupling$, that the drive tone is tuned to the upper (blue) mechanical sideband of the cavity: $\detune = - \mechfreq + \delta$, where $\delta \ll \mechfreq$.
After substitution of the detuning we apply the rotating wave approximation (RWA) which amounts to ignoring all the rapidly oscillating terms in the equation of motion~\eqref{eq:lang_env}.
As a result, the equations are greatly simplified.
In particular, we immediately see that the matrices $\mmat \Phi$ and $\mmat \Psi$ vanish as they are comprised of rapid terms only.
In the interesting case of driving exactly on the mechanical sideband~($\delta = 0$),
\begin{equation}
	\mmat \Theta ( - \mechfreq, \mechfreq ) \stackrel{\text{RWA}}{=} \bbsigma_1 =
	\begin{pmatrix}
		0 & 1
		\\
		1 & 0
	\end{pmatrix},
\end{equation}
so the drift and diffusion matrices take the simple time-independent form
\begin{equation}
	\mmat A\up{env}\s{RWA} =
	\begin{pmatrix}
		- \kappa \cdot \II_2 & g \cdot \bbsigma_1
		\\
		g \cdot \bbsigma_1 & - \tfrac \gamma 2 \cdot \II_2
	\end{pmatrix},
	\quad
	\mmat D\up{env}\s{RWA} = 2 \diag [ \kappa , \kappa , \Gamma , \Gamma ].
\end{equation}
These matrices can as well be used to compute a covariance matrix using a Lyapunov equation analogous to~\cref{eq:covmat_lyapunov}.

Importantly, since the coefficients in the equations of motion in RWA are time-independent, these equations can be solved analytically:
\begin{equation}
	\label{eq:sol_analytic}
	\mvec v\up{env} (t) = \me^{ \mmat A t } \mvec v\up{env} (0) + \int_0^t \dd \tau \me^{\mmat A (t - \tau) } \mvec \nu\up{env} (\tau).
\end{equation}
One can then proceed substituting this solution into the definition of the covariance matrix to compute the latter.

In the particular case of RWA an analytical solution for the quadratures can be obtained by substituting solution~\eqref{eq:sol_analytic} into the input-output relations~\eqref{eq:input_output} and the definition of the pulse quadratures~\eqref{eq:pulses_definition}.
Substitution of this solution into definition of the covariance matrix yields an analytical expression for $\mmat V$.

\subsection{Filtering functions}
\label{sec:filtering_functions}

In this section, we elaborate on the problem of choosing filtering functions $f\up{out}$ and illustrate certain choices thereof.
The problem associated with choosing these functions is usually considered in theory in the context of time-continuous measurement.
For recent reviews, see e.g.~\cite{chantasri_unifying_2021,lammers_quantum_2024}.

We start with the conceptually simplest regime when in addition to the condition of resolved sideband, the conditions of weak coupling $g \ll \kappa$, long pulses $\tau \gg \kappa^{-1}$, and good thermal isolation $\gamma,\Gamma \ll g,\kappa$ are satisfied.
In this regime, we can fully ignore the thermal environment of the mechanical oscillator, and moreover, adiabatically eliminate the intracavity optical mode.
The dynamics of the quadratures approaches a pure two-mode squeezing.
In more detail, the solutions of the Langevin equations read
\begin{align}
  \notag
  X\s{m} ( \tau ) & = \me^{ G \tau } X\s{m} (0)
  + \sqrt{ 2 G } \: \me^{ G \tau }  \int_0^\tau \dd{s} \me^{- G s} X\up{in} (s)
  \\
  & = \sqrt{ \ampgain } X\s{m} (0) + \sqrt{ \ampgain - 1 } \Mode X\up{in} (\tau),
  \\
  X\s{c} (t) & = \frac{ g }{ \kappa } X\s{m} (t) + \sqrt{ \frac{ 2 }{ \kappa }} X\up{in} (t).
\end{align}
Here we defined
\begin{gather}
  G = \frac{ g^2 }{\kappa },
  \quad
  \ampgain = \me^{ 2 G \tau },
  \\
  \Mode X\up{in} (t) = \sqrt{ \frac{ 2 G }{ 1 - \me^{ 2 G t }}} \int_0^t \dd{s} \me^{ - G s } X\up{in} (s).
\end{gather}
For the two other quadratures the solutions are obtained by trivial substitutions.
To derive an expression for the leaking light, we substitute the solution above into the standard input-output relations of the cavity:
\begin{multline}
  \label{eq:filter:solout}
  X\up{out}(t) = -  X\up{in}(t) + \sqrt{ 2 \kappa } X\s{c} (t)
  \\
  = \sqrt{ 2 G } \: \me^{ G t } X\s{m} (0) + 2 G \: \me^{ G t } \int_0^t \dd{s} \me^{ - G s } X\up{in} (s) + X\up{in} (t).
\end{multline}

To perform the bipartite optomechanical entanglement analysis, one has to define a certain mode of the leaking light, which is done by filtering the instantaneous values of the field with an envelope (as defined in~\cref{eq:pulses_definition}):
\begin{equation}
  \label{eq:filter:modedef}
  ( \Mode{X}\up{out},\Mode{Y}\up{out} ) = \int_0^\tau \dd{t}(X\up{out} (t),Y\up{out} (t)) f\up{out} (t).
\end{equation}
Let us remind that the mode profile function $f\up{out}(t)$ must be normalized in order for the mode quadratures to satisfy canonical commutation relations:
\begin{equation}
  \label{eq:filter:foutnorm}
  \int_0^\tau \dd{t} ( f\up{out} (t) )^2 = 1.
\end{equation}

Substitution of~\cref{eq:filter:solout} into the definition of the output mode quadratures~\cref{eq:filter:modedef} yields for $\Mode X\up{out}$
\begin{multline}
  \Mode X\up{out} = X\s{m} (0) \sqrt{ 2 G } \int_0^\tau \dd{t} \me^{ G t } f\up{out} (t)
  \\
  + \int_0^\tau \dd{t} X\up{in} (t)
  \left[ f\up{out} (t)
    +
    2 G \me^{ - G t } \int_t^\tau \dd{s} \me^{ G s } f\up{out} (s)
  \right]
  \\
  = X\s{m} (0) \sqrt{  \me^{ 2 G \tau } - 1  }\int_0^\tau \dd{t} f\up{ad}(t) f\up{out} (t)
  + \dots
\end{multline}
where $f\up{ad}(t) = \sqrt{ 2 G  / ( \me^{ 2 G \tau } - 1 )} \: \me^{ G t }$ satisfies the normalization~\eqref{eq:filter:foutnorm}.
The expression $\sqrt{ \eta } \equiv \int_0^\tau \dd{t} f\up{ad}(t) f\up{out} (t)$ can then be understood as a scalar product of these two functions.
As each of them has unit norm (defined by the scalar product), the maximal value $\sqrt \eta = 1$ is obtained by the choice $f\up{out} (t) = f\up{ad} (t)$.
With this choice, the input-output relations read~\cite{hofer_quantum_2011}
\begin{align}
	X\s{m} (\tau) & = \sqrt{ \ampgain } X\s{m} (0) + \sqrt{ \ampgain - 1 } \Mode X\up{in} (\tau),
	\\
	Y\s{m} (\tau) & = \sqrt{ \ampgain } Y\s{m} (0) - \sqrt{ \ampgain - 1 } \Mode Y\up{in} (\tau),
	\\
  \Mode X\up{out} & = \sqrt{ \ampgain} \Mode X\up{in} (\tau) + \sqrt{ \ampgain - 1 } X\s{m} (0),
	\\
  \Mode Y\up{out} & = \sqrt{ \ampgain} \Mode Y\up{in} (\tau) - \sqrt{ \ampgain - 1 } Y\s{m} (0).
\end{align}
These transformations correspond to an ideal unitary two-mode squeezing interaction between the optical and mechanical modes.

A different choice of the filtering function $f\up{out} (t) \neq f\up{ad} (t)$ ensures $\sqrt{\eta} < 1$.
Then, preservation of the commutation relations ensures that
\begin{multline}
  \label{eq:filter:wrong}
  \Mode X\up{out} = X\s{m} (0) \sqrt{ \ampgain - 1 } \sqrt{ \eta }
  \\
  + 
  \sqrt{ (\ampgain - 1) \eta + 1 } \int_0^\tau \dd{t} X\up{in} (t) h\up{in} (t),
\end{multline}
where $h\up{in}(t)$ is a different function that also satisfies normalization~\eqref{eq:filter:foutnorm}.
A direct computation of the two-mode squeezing shows that it is reduced compared to the case of $\eta = 1$.
This suggests that $f\up{out} = f\up{ad}$ is the optimal mode choice for the adiabatic limit.

In general case, where the assumptions of the adiabatic regime do not hold, the solution of the Langevin-Heisenberg equations becomes more complex~(see~\eqref{eq:sol_analytic}).
In analogy with the adiabatic regime, the coefficient at $X\s{m} (0)$ in the expression for $X_c(t)$ can be used as an initial guess for the filtering function.
This, however, turns out to be a suboptimal choice, as proven by the Bayesian optimization in the main text.

\subsection{Readout of the mechanical state}
\label{sec:readout_of_the_mechanical_state}

Verification of the optomechanical two-mode squeezing requires measuring both parties of the bipartite optomechanical system.
As a direct detection of the mechanical oscillator is impossible, indirect measurements of the mechanical state are usually performed using a subsequent optical pulse.
Importantly, for the purpose of verification, the only purpose of this pulse is an accurate read-out of the mechanical state.
Ideally, a verification pulse should not introduce any additional squeezing, but instead only passively capture the state of the mechanical oscillator after the first, squeezing, pulse.
In this section, we show how this operation is performed by a red-detuned pulse ($\Delta = \Omega\s{m}$).

Assuming that the verification stage takes place after a delay of the length $\tau\s{del}$, quadratures of the mechanical mode at the beginning of this pulse read
\begin{multline}
  \pma{ X\s{m} (\tau + \tau\s{del}) }{ Y\s{m} (\tau + \tau\s{del}) } = 
  \me^{ - \gamma \tau\s{del} / 2 }
  \pma{ X\s{m} (\tau) }{ Y\s{m} (\tau) }
  \\+
  \int_\tau^{\tau + \tau\s{del}} \dd{s}
  \me^{ - \gamma ( \tau\s{del} - s )}
  \pma{ X\s{th} (s) }{ Y\s{th} (s) }.
\end{multline}
The only process that takes place during the delay is the thermalization of the mechanical oscillator to the environment, which only passively attenuates its quadratures and admixes thermal noise.

Starting for simplicity with the adiabatic regime, we can write the transformations for the mechanical mode and the intracavity light mode during the verification pulse as (for brevity, we reset notation for the instant of the beginning of the verification pulse to zero $t = \tau + \tau\s{del} \mapsto t = 0$)
\begin{align}
  \notag
  X\s{m} (  t ) & = \me^{ - G t } X\s{m} (0 )
  + \sqrt{ 2 G } \: \me^{ - G t }  \int_0 ^{ t} \dd{s} \me^{G s} X\up{in} (s),
  \\
  X\s{c} ( t) & = \frac{ g }{ \kappa } X\s{m} ( t) + \sqrt{ \frac{ 2 }{ \kappa }} X\up{in} ( t),
\end{align}
Consequently, for the leaking field, one can write
\begin{multline}
  X\up{out}\s{RO}(t) = -  X\up{in}(t) + \sqrt{ 2 \kappa } X\s{c} (t)
  \\
  = \sqrt{ 2 G } \: \me^{ - G t } X\s{m} (0) + 2 G \: \me^{ -  G t } \int_0^t \dd{s} \me^{ G s } X\up{in} (s) + X\up{in} (t),
\end{multline}
and
\begin{multline}
  \Mode X\up{out}\s{RO} 
  = X\s{m} (0) \sqrt{ 1 -  \me^{ - 2 G \tau }}\int_0^\tau \dd{t} f\up{ad}\s{RO}(t) f\up{out} (t)
  + \dots
  \\
  = X\s{m} (0) \sqrt{ 1 - \mathfrak T} \sqrt{ \eta } + \dots .
\end{multline}
Where, again for brevity, we assume the length of the verification pulse also $\tau$, and
\begin{equation}
  f\up{ad}\s{RO} (t) = \sqrt{ \frac{ 2 G }{ 1 - \me^{ - 2 G \tau }}} \me^{  - 2 G t  },
  \quad
  \mathfrak T = \me^{ - 2 G \tau }.
\end{equation}
In the case of the verification pulse, the optimal choice of the filtering function is straightforwardly $f\up{out} = f\up{ad}\s{RO}$, as the only figure of merit for this pulse is the transfer of the mechanical quadratures to light.
We see that for sufficiently long $\tau$, we can achieve a perfect transfer $\Mode X\up{out}\s{RO} = X\s{m} (0)$.

This logic, however, applies equivalently well to the non-adiabatic regime, where the solution for the instantaneous leaking light reads
\begin{equation}
  X\up{out}\s{RO} (t) = X\s{m} (0) h\up{m} (t) + \dots.
\end{equation}
The filtering function that would maximize the contribution of $X\s{m}(0)$ in $\Mode X\up{out}\s{RO}$ is (up to normalization) $h\up{m}$.
Any different choice of the filtering function would decrease this contribution.

\section{Simulating optimizations}
\label{app:simulation}
\subsection{Simulation of the optomechanics}
\label{sec:sim_of_optomech}
The results presented in~\secref{sec:OptimResults} are produced using a
simulation of the optomechanical system under study developed in Python.
The initial value problem solver in SciPy (\pythoninline{scipy.integrate.solve_ivp}) 
is used to compute the covariance matrix $\covmat$ by solving~\eqref{eq:covmat_lyapunov},
with the generalized two-mode squeezing $\gensqz$ given by~\eqref{eq:GenSqz}.
An analytical solution to $\covmat$ for the constant (rectangular) pulse, 
utilizing a rotating wave approximation (RWA),
is used later in this appendix for comparison with the numerical solution (which does not require the RWA).
The analytical solution is described in Ref.~\cite{Rakhubovsky2018}.
%-----------------------------------------------------------------------------------

\subsection{Optimization of parameters}
\label{sec:optim_of_params}
The Gaussian process minimization function (\pythoninline{skopt.gp_minimize})
from the Scikit Optimize library \cite{head_tim_2021_5565057}
is used for Bayesian optimization.
The initial Gaussian process model is built from evaluations of the cost function,
for instance $-\gensqz$, based on pseudo-random samples of the optimization variables.
An \textit{acquisition function} determines the choice of variables for each optimization step.
The default acquisition method \pythoninline{gp_hedge} automates a combination of \textit{exploration} and \textit{exploitation} steps,
with each step requiring one evaluation of the cost function.

The number of initial samples $N_{\rm initial}$ and optimization steps $N_{\rm optim}$ 
required for a reliable outcome increase with the number of optimization variables
(the dimensionality of the optimization search space $N_{\rm d}$).
The choice of values also depends on the topology of the cost function and so they are difficult to estimate.
The values used in this study are summarized in \tblref{tbl:bo_steps}.
For the lower dimensional spaces, similar reliability is found with far fewer steps than those listed,
but not so for the highest dimensional space.

\begin{table}
    \caption{
        \textbf{Bayesian optimization steps.}
        The number of initial samples $N_{\rm initial}$ and optimization steps $N_{\rm optim}$ used for the specific optimization variable combinations.
        The dimensionality of the search space $N_{\rm d}$ is equivalent to the number of optimization variables.
        The time-dependent functions $\couplingtd, \measfunctd, \detuneofftd$ are piecewise linear parameterized by 6 variables each.
    }
    \label{tbl:bo_steps}
    \begin{center}
        \renewcommand{\arraystretch}{1.3}
        \setlength\tabcolsep{8pt}
        \begin{tabular}{l r r r}
            \thickhline
            variables & $N_{\rm d}$ & $N_{\rm initial}$ & $N_{\rm optim}$ \\
            \hline
            $\coupling, \pulsedur$ & 2 & 100 & 400\\
            $\couplingtd, \pulsedur, \measfunctd$ & 13 & 200 & 600\\
            $\measfunctd$ & 6 & 100 & 400\\
            $\couplingtd, \pulsedur, \measfunctd, \detuneofftd$ & 19 & 200 & 600\\
            \thickhline
        \end{tabular}
    \end{center}
\end{table}

The Gaussian process minimization is computationally expensive,
and so a HPC cluster was used to process the repeated optimizations in parallel,
utilising the GNU Parallel library~\cite{GnuParallel}.
For some optimizations (see \secref{sec:detect_optim}) it is sufficient to use a gradient-based algorithm.
Where first and second order derivatives are available,
a conjugate gradient method is used, 
specifically the `Newton conjugate gradient' method in \pythoninline{scipy.optimize}.
Where the derivatives need to be estimated, the 
L-BFGS-B implementation in SciPy is used~\cite{Byrd1995}.
%-----------------------------------------------------------------------------------

\subsection{Maximizing logarithm negativity}
\label{sec:logneg_optim}
\begin{figure}
	\centering
	\includegraphics[width=0.95\linewidth]{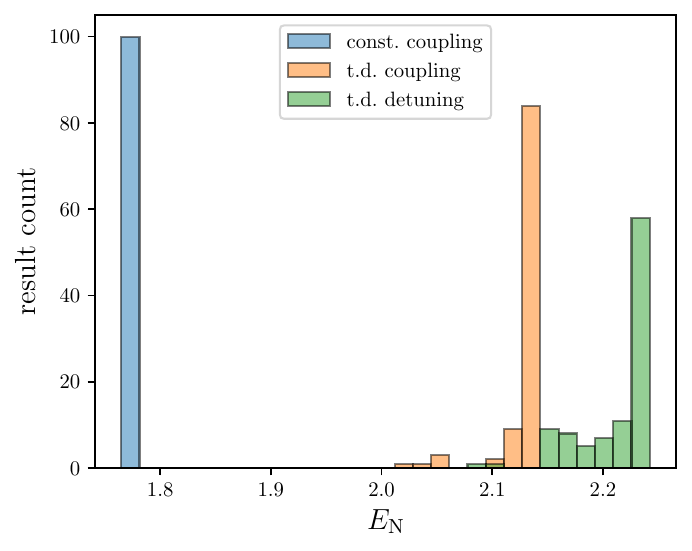}

	\caption{\textbf{Optimized logarithmic negativity.}
		Results of optimizing pulse parameters to maximize logarithmic negativity $\logneg$.
		The distribution of 100 repetitions are shown
		for each of the 3 different pulse parameterization schemes.
		The dataset labels refer to the same parameterization schemes as in \figref{fig:optimal_gen_sqz}.
	}
	\label{fig:logneg_optim}
	% plotted using: plot_result_hist.py
	% data:
	% 'const. coupling': ../data/theo_sys-const_optimal-numeric-cool_osc-log_neg-sko-hc-optim_results.7448062-combined.txt
	% 't.d. coupling': ../data/theo_sys-5slice_pwl-cool_osc-logneg-sko-optim_results.7447938-combined.txt
	% 'detuning': ../data/theo_sys-5slice_pwl-td_detune-cool_osc-log_neg-sko-optim_results.7591523-combined.txt
	% with param files:
	% ../parameters/ncl-params-theo_sys-const_optimal-numeric-cool_osc-log_neg-sko-hc.ini
	% ../parameters/ncl-params-theo_sys-5slice_pwl-cool_osc-log_neg-sko.ini
	% ../parameters/ncl-params-theo_sys-5slice_pwl-td_detune-cool_osc-log_neg-sko.ini
	
\end{figure}

Logarithmic negativity $\logneg$ and generalized two-mode squeezing $\gensqz$ 
are both measures of nonclassicality (see \appxref{app:measures}).
However, they are not equivalent, and do not strongly correlate, 
hence optimizing pulse parameters to maximize $\gensqz$ does not necessarily also maximize $\logneg$.

The results for specifically optimizing pulse parameters to maximize $\logneg$ are shown in \figref{fig:logneg_optim}.
We see the same trend in the maximum achievable value as with maximizing $\gensqz$ (seen in \figref{fig:optimal_gen_sqz}):
modulating the coupling strength provides improvement, and further improvement again with modulation of the detuning.
We also see similar reliability in the optimization process, 
with $> 50\%$ of attempts succeeding in finding close to the global maximum.

%-----------------------------------------------------------------------------------

\subsection{Local traps in optimization landscape}
\label{sec:local_traps}
\begin{figure}
	\centering
	\begin{tikzpicture}
	\node(a){\includegraphics[width=0.95\linewidth]{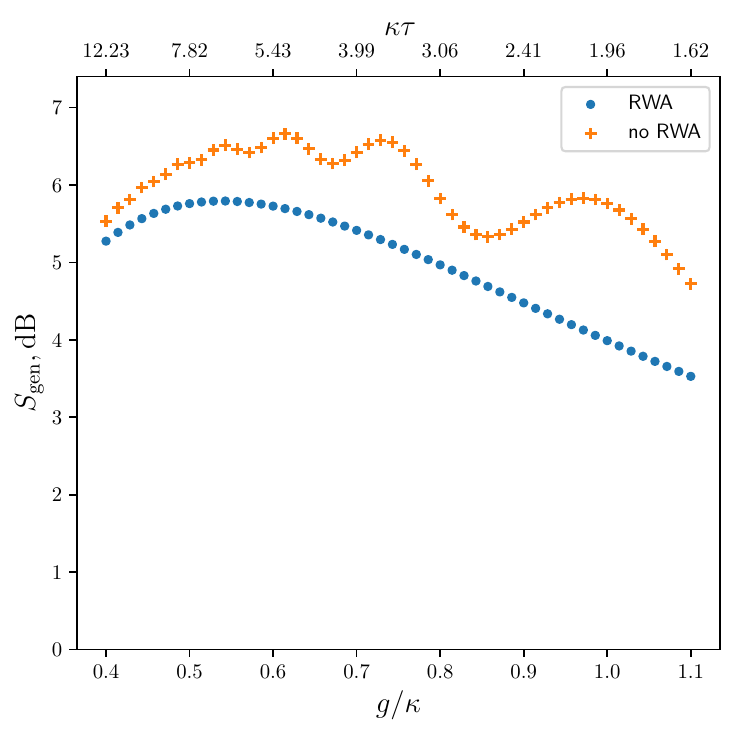}};
	\node at (a.south west)
	[
	anchor=center,
	xshift=39.5mm,
	yshift=30mm
	]
	{
		\includegraphics[width=0.58\linewidth]{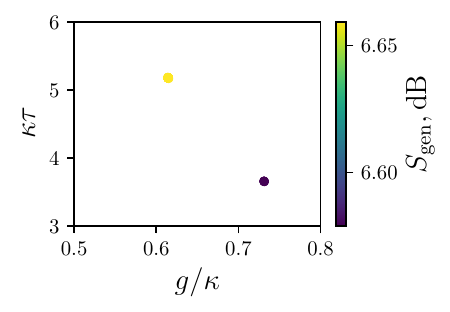}
	};
	\end{tikzpicture}

	\caption{\textbf{Optimal coupling and pulse duration for a constant coupling strength pulse.}
		The main plot shows the generalized squeezing $\gensqz$ for different coupling strengths $\coupling$.
		The value of $\pulsedur$ is chosen is such that the amplitude gain is at its upper limit 
		$\ampgain = \exp{\left(2 \coupling^2 \pulsedur \right)} = \ampgainlimit$.
		The value of $\gensqz$ differs depending on whether the rotating wave approximation (RWA)
		is used in the computation of the covariance matrix $\covmat$.
		The blue dots indicate $\gensqz$ computed analytically using the RWA 
		and the orange crosses show $\gensqz$ computed by the numerical solver.
		The inset shows the values $\coupling,\pulsedur$ resulting from
		100 repetitions of optimizing for maximal $\gensqz$ (without the RWA),
		with $N_{\rm initial}=50, N_{\rm optim} = 100$.
		The grouping of points correspond with the near-degenerate maxima for $\gensqz$ in the main plot.
	}
	\label{fig:const_coup_dur_values}
	% Main
	% Plot generated by: compare_exact_numeric.py
	% ../parameters/ncl-params-theo_sys-const_optimal-numeric-cool_osc-compare_rwa_plt.ini
	% Inset
	% plotted using: plot_result_scatter.py
	% data:
	% ../data/theo_sys-const_optimal-numeric-cool_osc-sgen-sko-optim_results.7440046-combined.txt

\end{figure}

The optimization process should ideally result
in the same optimal value for the figure of merit with each repeated run.
That is, the maximum possible value of $\gensqz$ should be returned reliably within some acceptable tolerance.
In~\secref{sec:OptimResults} we see that this is not always the case,
especially for the higher dimensional search spaces.
Failure to reliably find the global maximum implies the algorithm has found a local maximum or \emph{trap}.

We see in \figref{fig:const_coup_dur_values} that for the constant pulse
there are near degenerate solutions
at $\coupling \approx 0.61 \optdamp$ and $\coupling \approx 0.73 \optdamp$.
In \figref{fig:optimal_gen_sqz} we see that the Bayesian optimization algorithm,
with sufficient steps, finds the global maximum for $\gensqz$ in approximately 98\% of attempts.
However, with fewer steps, 
the algorithm more frequently finds the slightly lower value of $\gensqz$ at $\coupling \approx 0.73 \optdamp$.

For a high frequency oscillator, $\mechfreq > 50 \optdamp$, 
the numeric solution converges to the analytic (RWA) solution,
where there are no potential traps in the optimization space.
From this we can speculate that these undulations are related to resonance effects.
We can assume that there also traps in the higher dimensional search spaces,
at least in the PWL parameters of $\couplingtd$,
which are more difficult for the algorithm to avoid due to the vastness of the space.
Consequently, we see lower reliability (greater variance in $\gensqz$) for the `t.d.~detuning' result set in \figref{fig:optimal_gen_sqz}.
However, when the algorithm does find the global maximum,
we see in \figref{fig:detuning_pulses} that the solution for $\couplingtd, \measfunctd, \measfunctd$ is quite distinct.

\begin{figure}
	\centering
	\includegraphics[width=0.9\linewidth]{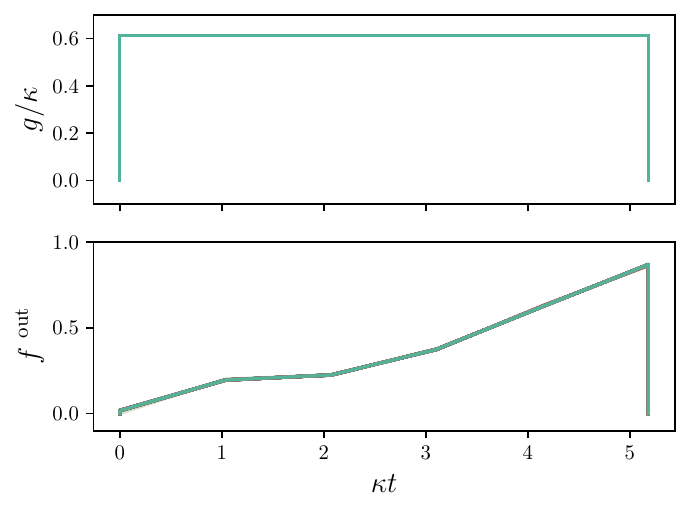}
	\caption{\textbf{Optimal measurement function profiles with fixed constant coupling.}
		The time-dependent profiles for the optimal measurement function $\measfunctd$
		resulting from the 100 to optimization attempts.
		These correspond to the results shown in~\figref{fig:optimal_gen_sqz}, `$\measfunc$ only' dataset.
		The coupling $\coupling$ and pulse duration $\pulsedur$ are fixed at the optimal values
		found previously, and the only variables are the PWL parameters of $\measfunctd$.
		All 100 pulses are close to identical, which makes them almost indistinguishable in this figure.
	}
	\label{fig:optimal_pulse_fout}
	% plotted using: plot_pulse.py
	% data:
	% ../data/theo_sys-5slice_pwl-fout_only-optim_results.185304-combined.txt
	% with param files:
	% ../parameters/ncl-params-theo_sys-5slice_pwl-fout_only.ini
\end{figure}

We see evidence in~\figref{fig:optimal_gen_sqz}
(`$\measfunc$ only' dataset) to suggest that traps are less prevalent in the measurement function dimensions.
When optimizing only the measurement function $\measfunctd$ in repeated attempts,
the maximal value has a very narrow distribution,
reliably finding $\gensqz=6.78\,\decibel$.
Correspondingly we find a distinct solution for $\measfunctd$, shown in \figref{fig:optimal_pulse_fout}.

%-----------------------------------------------------------------------------------

\subsection{High temperature bath}
\label{sec:hot_bath_num_stab}
\begin{figure}
	\centering
	\includegraphics[width=0.99\linewidth]{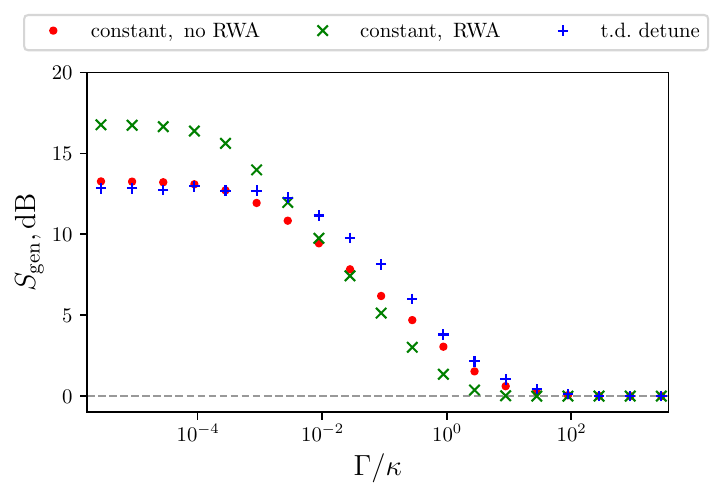}
	\caption{\textbf{Achievable generalized two-mode squeezing over thermal range comparing coupling pulse variants.}
		The two-mode generalized squeezing $\gensqz$ found through optimization
		(maximum from 10 repetitions)
		of the driving pulse parameters
		are shown for the given reheating rate $\reheatrate/\optdamp$.
		The red dots show the squeezing achieved through optimization of a constant strength pulse,
		with the covariance matrix $\covmat$ computed without using the rotating wave approximation (RWA).
		The green diagonal crosses are also from optimization of a constant pulse,
		but using the analytic computation of the covariance matrix, which requires the RWA.
		The blue upright crosses show the result of optimizing the piecewise linear  parameters
		of the coupling strength $\couplingtd$, measurement function $\measfunctd$ and detuning $\detunetd$.
	}
	\label{fig:compare_optim_thermal_range}
	% data generated using: srun-ncl-n_ph_range-optim-sko-con-red_rng.sh;srun-ncl-n_ph_range-optim-sko-coe-red_rng.sh;srun-multi_run_optim-skopt-tdd-n_ph.sh
	% plotted using: plot_n_ph_sqz.plot_sqz_thermal_range
	% data in:
	% ../data/n_ph_range-theo_sys-const_optimal-numeric-cool_osc-sgen-sko-red_range-try1
	% ../data/n_ph_range-theo_sys-const_optimal-exact-cool_osc-sgen-sko-red_range-try1
	% ../data/n_ph_range-theo_sys-td_detune-cool_osc-sgen-sko
	% with param files:
	% (main fig)
	% ../parameters/ncl-params-theo_sys-cool_osc-sqz_rng_plt-red_rng-sko.ini
	% data generated with param files:
	% ../parameters/ncl-params-theo_sys-const_optimal-numeric-cool_osc-sgen-sko.ini
	% ../parameters/ncl-params-theo_sys-const_optimal-exact-cool_osc-sgen-sko.ini
    % ../parameters/ncl-params-theo_sys-5slice_pwl-td_detune-cool_osc-sgen-sko.ini
\end{figure}

The two-mode generalized squeezing $\gensqz$ achievable through optimization of the driving pulse
parameters is shown in \figref{fig:compare_optim_thermal_range} for a set of reheating rates $\reheatrate$
arising from bath temperatures in the range  $10^{4} < \ntherm < 10^{13}$ with fixed mechanical damping $\mechdamp = 2.8 \times 10^{-10}$.
We see that modulating the coupling strength and detuning allows for greater squeezing for all reheating rates $\reheatrate > 10^{-3}\optdamp$.
However, there is no strong evidence here that modulating the pulse extends the reheating range at which measurable squeezing can be achieved.
The limit for this is found to be $\reheatrate=88.5\optdamp$, with $\gensqz \approx 0.055\,\decibel$ for the constant pulse
and $\gensqz \approx 0.120\,\decibel$ for the modulated pulse.
For $\reheatrate \ge 280\optdamp$, squeezing is below the measurable limit $\gensqz = 0.05\,\decibel$ for both pulse types.

In the low heating range, we see that the constant strength pulse appears to out-perform the modulated pulse.
This is an artefact related to the difficultly that the optimization algorithm has in navigating this parameter space.
The parameters equivalent to constant pulse are accessible to the algorithm when optimizing $\couplingtd, \measfunctd,\detunetd$,
but the landscape flattens at the extremes of the coupling,
and the solution tends to the upper bound $\coupling = 2.0 \optdamp$ at these low temperatures,
making it more difficult for the algorithm to reach the optimal solution.

There is a significant divergence of the squeezing predicted by
the numerical and analytic solving methods for the constant pulse
in the both the low and high reheating range.
The computations of the covariance matrix $\covmat$ differ due to the necessary use of the RWA to calculate an analytic solution.
The numerical and analytic solution squeezing values converge for fast oscillations $\mechfreq \gtrsim 50 \optdamp$.
The long tail of measurable squeezing at high temperatures
is an interesting and unexpected feature
that only appears for low frequency oscillations $\mechfreq \approx \optdamp$.
It could potentially be exploited to obtain measurable squeezing if extremely high coupling strength can be achieved.

\begin{figure}[!ht]
	\centering
	\includegraphics[width=0.99\linewidth]{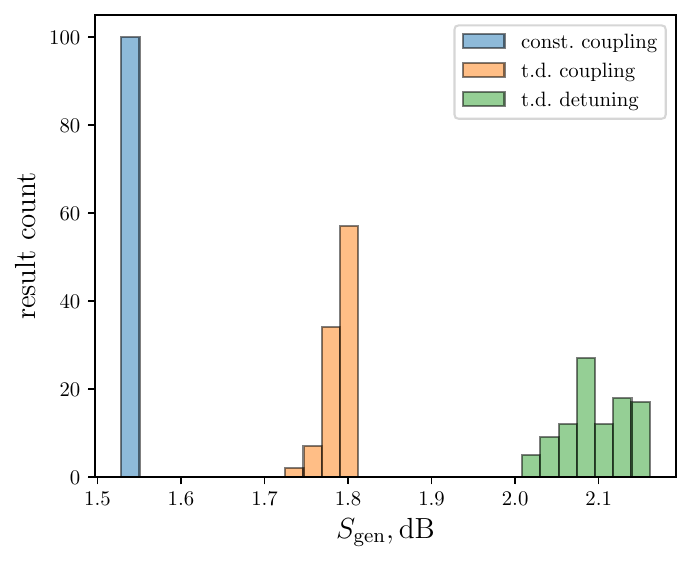}
	\caption{\textbf{Optimized generalized squeezing at high bath temperature.}
		Results of optimizing pulse parameters to maximize general squeezing for a bath temperature
		resulting in high reheating rate $\reheatrate=2.8\optdamp$.
		The distribution of 100 repetitions for each of the 3 different pulse parameterization schemes are shown.
		The dataset labels refer to the same parameterization schemes as in  \figref{fig:optimal_gen_sqz}.
	}
	\label{fig:optimal_gen_sqz_coolish_osc_hottish_bath}
	% plotted using: plot_result_hist.py
	% data:
	% 'const. coupling': ../data/theo_sys-const_optimal-numeric-cool_osc-hotish_bath-sgen-sko-hc-optim_results.7607241-combined.txt
	% 't.d. coupling': ../data/theo_sys-5slice_pwl-cool_osc-hotish_bath-sgen-sko-optim_results-7607242+7607457-combined.txt
	% 't.d. detuning': ../data/theo_sys-5slice_pwl-td_detune-cool_osc-hotish_bath-sgen-sko-optim_results.7607243+7607456-combined.txt
	% data generated with param files:
	% ../parameters/ncl-params-theo_sys-const_optimal-numeric-cool_osc-hotish_bath-sgen-sko-hc.ini
	% ../parameters/ncl-params-theo_sys-5slice_pwl-cool_osc-hotish_bath-sgen-sko.ini
	% ../parameters/ncl-params-theo_sys-5slice_pwl-td_detune-cool_osc-hotish_bath-sgen-sko.ini
\end{figure}

The distribution of repeated optimization results with a high temperature bath ($\reheatrate=2.8\optdamp$)
results are presented in \figref{fig:optimal_gen_sqz_coolish_osc_hottish_bath}.
The squeezing levels are much lower at this bath temperature, as expected.
We see the same relationship between different pulse types as with reheating rate $\reheatrate=0.063\optdamp$ in \figref{fig:optimal_gen_sqz}.
That is, the achievable squeezing increases by allowing modulation of the coupling strength and further again through modulation of the detuning.
%-----------------------------------------------------------------------------------

\subsection{Optimization of detection angles}
\label{sec:detect_optim}

\begin{figure}
	\centering
	\begin{minipage}{\linewidth}
		\centering
		a.) Thermal equilibrium oscillator
		\includegraphics[width=0.9\textwidth]{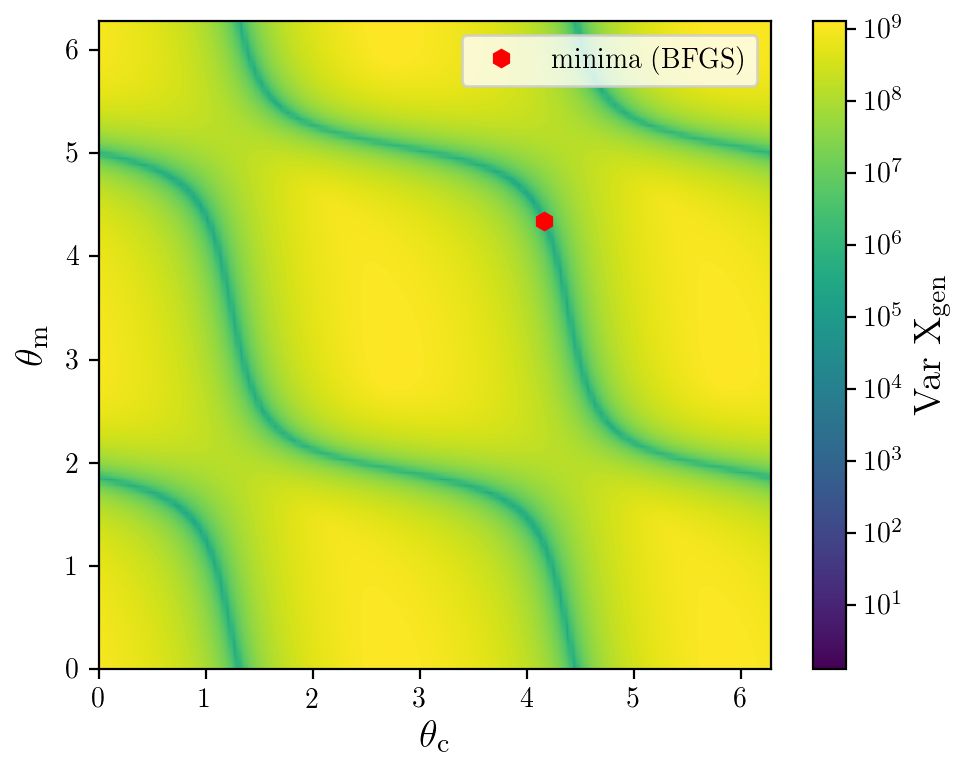}
	\end{minipage}
	\begin{minipage}{\linewidth}
		\centering
		b.) Cooled oscillator
		\includegraphics[width=0.9\textwidth]{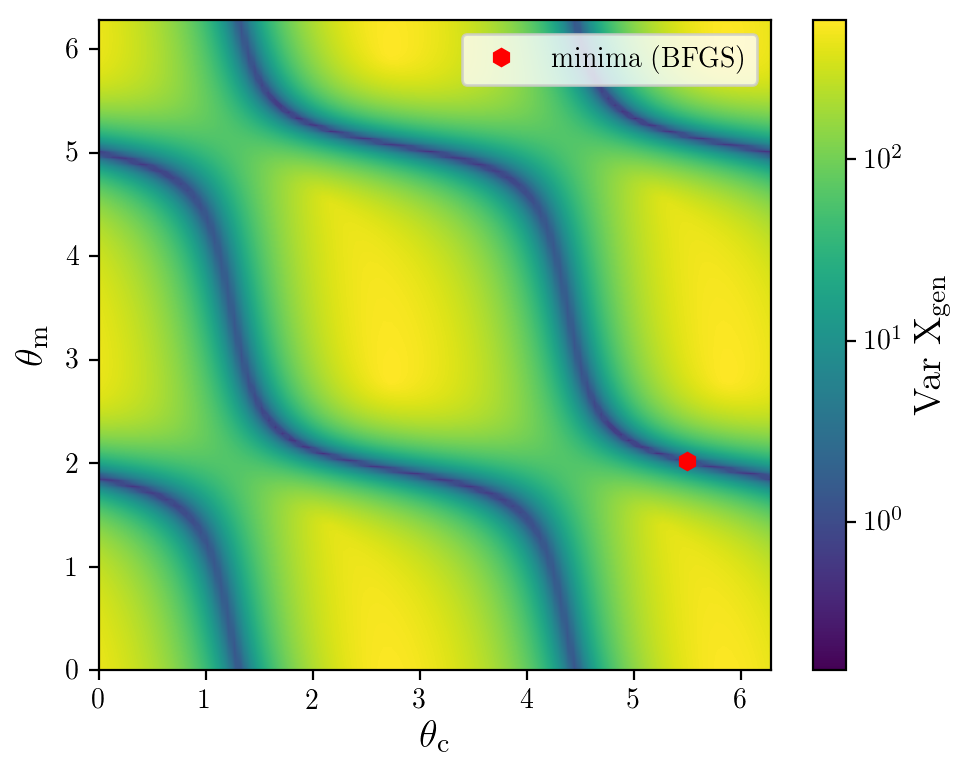}
	\end{minipage}
	\caption{\textbf{The squeezing indicator $\varxgen$ for detection angle ranges}.
		The contour plots represent the values $\varxgen\left[\qangc,\qangm\right]$
		for the covariance matrix generated by the optimal parameters of 
		$\couplingtd, \measfunctd, \detunetd, \pulsedur$, as described in \secref{sec:detuning}.
		The lighter (yellow) areas are where $\varxgen$ is greatest.
		The minimal value, which is equivalent to $\mineig$, is found in the dark (blue) trenches.
		In panel a.) the oscillator is initially in thermal equilibrium $\nmech=\ntherm$
		and in b.) the oscillator is initially cooled to $\nmech=100$.
		}
	\label{fig:min_eig_for_detect_angles}
	% originally plotted using: plot_ctrlparamland.plot_detect_angle_landscape
	% but this is now moved
	% To produce similar use: detection.plot_detect_angle_landscape via test_min_eigval_detect
	%NOTE: These are PNG images. There is no point to make them PDF, just results in much greater main PDF file size.
	% a.) is produced using the ncl-params-theo_sys-const_optimal-exact.ini params
	% b.) and c.) from ncl-params-theo_sys-5slice_pwl-td_detune.ini using the best parameters found through optimization.
\end{figure}

In \secref{sec:detect} we saw that numerical optimization can be used to determine the homodyne measurement angles $\qangc, \qangm$
that would allow direct measurement of $\varxgen$ equivalent to $\mineig$,
and hence calculate the two-mode squeezing $\gensqz$.
The Newton conjugate gradient method is used to determine the optimal weighting $\qangw$
without the need for additional computations of the covariance matrix $\covmat$.
Hence the optimization space is reduced to two dimensions,
$\qangc$ and  $\qangm$.
The optimization landscape of $\varxgen\left[\qangc,\qangm\right]$
can be seen in the contour plots of~\figref{fig:min_eig_for_detect_angles}.
The L-BFGS-B algorithm is used to find the optimal $\qangc, \qangm$ that minimize $\varxgen$.

The form of~\eqref{eq:gentwomode} implies that 
$\varxgen\left[\qangc,\qangm\right]$ is periodic in $\qangc, \qangm$.
The physical interpretation implies that the period should be $\pi$.
The periodicity visible in~\figref{fig:min_eig_for_detect_angles} confirms this physical interpretation.

When the mechanical oscillator is initially in thermal equilibrium with the environment
($\nmech = \ntherm = 2.26 \times 10^8$),
the range of $\varxgen\left[\qangc,\qangm\right], 0 < \qangc,\qangm < 2\pi$, is very large,
with some values of $\varxgen > 10^9$.
As the minimum value of $\varxgen$ is less than 1, 
and coupled with the non-uniformity of the landscape --
the dark trenches are not uniformly deep, with minima at specific locations --
the algorithm finds it difficult to navigate to the optimal solution.
With the initially cooled oscillator ($\nmech = 100$) the range is much reduced,
and so the algorithm reliably finds the minimal value of $\varxgen$
that is equivalent to $\mineig$.
Approximately 200 function evaluations gives a solution to satisfactory precision.
%-----------------------------------------------------------------------------------

\subsection{Optimization with noisy controls}
\label{sec:noisy_optim}
\begin{figure}
	\centering
	\begin{minipage}{\linewidth}
		\centering
		\includegraphics[width=0.99\textwidth]{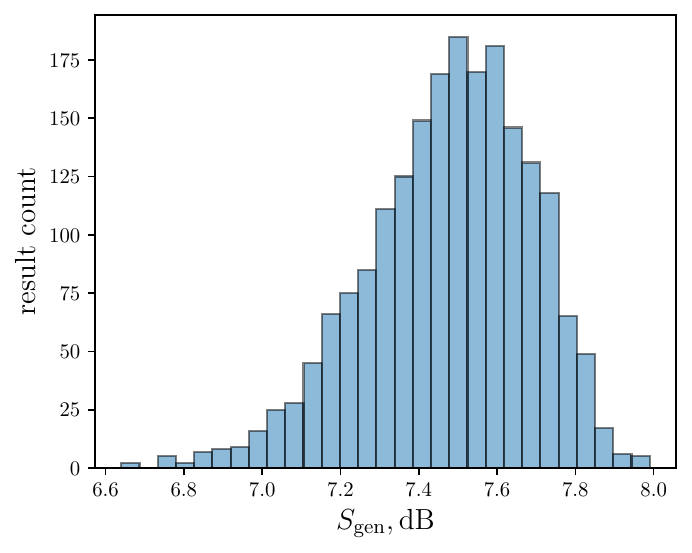}
	\end{minipage}
	\caption{\textbf{Generalized squeezing for optimal coupling pulse parameters with added noise.}
		The histogram illustrates the distribution of generalized two-mode squeezing $\gensqz$
		when noise (Gaussian with standard deviation of $10\%$)
		is added to the piecewise linear parameters, $\coupling_i$, of the coupling strength $\couplingtd$.
		The specific mean parameters for $\coupling_i$ are those found when optimizing
		with control noise included the simulation (see~\secref{sec:noise}).
		The histogram is generated from 2000 computations of $\gensqz$ with noisy parameters.
	}

	\label{fig:noisy_opt_pwl}
	% The data was produced by running: run_compute_covmat.py
	% plotted using: plot_result_hist.py
	% data:
	% ../data/ncl-theo_sys-5slice_pwl-cool_osc-noisy-many_rep-covmat_attribs.txt
	% with param file:
	% ../ncl-params-theo_sys-5slice_pwl-cool_osc-noisy-repeat.ini
\end{figure}

To simulate the potential effects of control noise on the optimization process,
the piecewise linear parameters, $\coupling_i$, of $\couplingtd$ are modified 
from those proposed by the optimization algorithm when calculating the figure of merit (FOM) that is used to update the algorithm's model.
These values of $\coupling_i^{\rm (FOM)}$ for the FOM are sampled from a Gaussian distribution with mean $\coupling_i$ and standard deviation $\coupling_i / 10$.
The values are truncated such that $\coupling_i^{\rm (FOM)} \ge 0$ and are no more than 3 standard deviations from the proposed value.
Note that the amplitude gain limit is not maintained, and hence potentially $\ampgain > \ampgainlimit$ for the pulse used to compute the FOM,
therefore values for $\gensqz$ may exceed those possible when noise is not added.
The final maximum squeezing value found through optimization (reported in \secref{sec:noise}) is calculated without noise,
otherwise the noise may obscure the true result.

The effects of the noise on $\gensqz$ can be seen in \figref{fig:noisy_opt_pwl}.
The distribution of $\gensqz$ computed with noise added to the optimal pulse
resembles a skewed Gaussian, with a peak at $\gensqz \approx 7.5\,\decibel$.
This peak corresponds with the maximum value of $\gensqz$ found when optimizing $\couplingtd, \measfunctd, \pulsedur$,
as we would expect.

\bibliography{tmso_allrefs}

\end{document}